\documentclass[
 reprint,
 amsmath,amssymb,
 aps
]{revtex4-2}

\usepackage{graphicx}
\usepackage{dcolumn}
\usepackage{bm}
\usepackage[normalem]{ulem}
\usepackage{xcolor}
\usepackage{newtxtext}
\usepackage{lmodern,pdftexcmds}
\usepackage{xstring}
\usepackage{mathrsfs}
\usepackage{soul}

\DeclareMathAlphabet{\mathbfsf}{\encodingdefault}{\sfdefault}{bx}{sl}
\renewcommand{\Vec}[1]{\bm{#1}}
\def\Tens#1{\IfSubStr{ABCDEFGHIJKLMNOPQRSTUVWXYZabcdefghijklmnopqrstuvwxyz}{#1}{\mathbfsf{#1}}{\bm{#1}}}
\newcommand{\figref}[1]{Fig.\,\ref{#1}}
\newcommand{\Figref}[1]{Figure\,\ref{#1}}
\newcommand{\figrefs}[2]{Figs.\,\ref{#1} and \ref{#2}}
\newcommand{\Figrefs}[2]{Figures\,\ref{#1} and \ref{#2}}
\newcommand{\figrefp}[2]{Fig.\,\ref{#1}\,(#2)}
\newcommand{\Figrefp}[2]{Figure\,\ref{#1}\,(#2)}
\newcommand{\Appendref}[1]{Appendix\,\ref{#1}}

\definecolor{green}{rgb}{0,0.5,0}

\begin{document}

\preprint{APS/123-QED}

\title{Emergent Stripes of Active Rotors in Shear Flows}

\author{Zhiyuan \surname{Zhao}%
$^{1,2}$}%
\author{Boyi \surname{Wang}$^{2,3}$}%
\author{Shigeyuki \surname{Komura}$^{1,4,5}$}%
\author{Mingcheng \surname{Yang}$^{2,3,6}$}%
\author{Fangfu \surname{Ye}$^{1,2,3,4,6}$}%
\author{Ryohei \surname{Seto}$^{1,4,7}$}%
\email{seto@ucas.ac.cn}%
\affiliation{%
$^1$Wenzhou Institute, University of Chinese Academy of Sciences, Wenzhou, Zhejiang 325000, China}%
\affiliation{%
$^2$School of Physical Sciences, University of Chinese Academy of Sciences, Beijing 100049, China}
\affiliation{%
$^3$Beijing National Laboratory for Condensed Matter Physics and Laboratory of Soft Matter Physics, Institute of Physics, Chinese Academy of Sciences, Beijing 100190, China}%
\affiliation{%
$^4$Oujiang Laboratory (Zhejiang Lab for Rengerative Medicine, Vision and Brain Health), Wenzhou, Zhejiang 325000, China}%
\affiliation{%
$^5$Department of Chemistry, Graduate School of Science, 
Tokyo Metropolitan University, Tokyo 192-0397, Japan}%
\affiliation{%
$^6$Songshan Lake Materials Laboratory, Dongguan, Guangdong 523808, China}
\affiliation{%
$^7$Graduate School of Information Science, University of Hyogo, Kobe, Hyogo 650-0047, Japan}

\date{\today}


\begin{abstract}
The shear-induced self-organization of active rotors into stripy aggregates is studied by carrying out computational simulations. 
The rotors, modeled by monolayers of frictional spheres, 
develop to stripy microstructures only when they counterrotate with respect to the vorticity of the imposed shear flow.
The average width of the stripes is demonstrated to be linearly dependent on 
the relative intensity of active torque to the shear rate. 
By giving insight into three collective particle behaviors, 
i.e., shear-induced diffusion, rotation-induced rearrangement, and edge flows,
we explain the mechanisms of formation of the particle stripes. 
Additionally, the rheological result shows the dependence of shear and rotational viscosities on the active torque direction and the oddness of the normal stress response. 
By exhibiting a collective phenomenon of active rotors, 
our study paves the way to understanding chiral active matter. 
\end{abstract}

\maketitle


\section{Introduction}
Nonequilibrium collective motion is one of the most intriguing behaviors of active matter\,\cite{marchetti2013hydrodynamics,zottl2016emergent,snezhko2016}.
For a collection of rotating active matter, such as spinning biological organisms\,\cite{riedel2005,drescher2009,petroff2015fast} 
and artificial rotors driven by external fields%
\,\cite{friese1998,grzybowski2000,tsai2005chiral,aubret2018targeted,sabrina2018shape,yang2021topologically}, 
their rotational motion can be transferred to translations through interactive hydrodynamic and/or contact forces, 
resulting in spontaneous self-organization that favors rotation in the same direction\,\cite{nguyen2014,yeo2015}.
In the last decade, intensive experimental and numerical works have studied the phase separation of counterrotating active matter\,\cite{nguyen2014,yeo2015,scholz2018rotating,shen2020hydrodynamic} and self-organization of corotating active matter into dynamic crystalline clusters\,\cite{goto2015,yan2015,petroff2015fast,singh2016universal,oppenheimer2019rotating}, 
in both quiescent liquids and a dry environment.
By applying wall-confined shear flows,
prior work\,\cite{yeo2010rheology} has reported that the rotating particles can hydrodynamically interact with the walls 
and self-organize into hexagonally structured strings. 
Besides the pattern constructions\,\cite{yeo2010rheology,shen2020two,driscoll2017unstable}, 
the existence of boundaries, such as impermeable walls or interfaces,
also causes asymmetric responses and then edge flows of the rotating active matter\,\cite{van2016spatiotemporal,scholz2018rotating,soni2019odd,yang2020,liu2020oscillating}.
Studying the features and mechanisms of collective phenomena of rotating active matter can not only further the understanding of chiral active systems, 
but also contribute to developing potential applications and experimental schemes.


In this paper, we show with computational simulations that a monolayer of spherical rotors, 
which counterrotate with respect to the vorticity of applied simple shear flow, 
can self-organize to stripelike aggregates that are featured with edge flows.
Besides the hydrodynamic lubrications, 
the simulations also take into account the frictional contacts between the particles, 
which have been demonstrated to play a significant role in developing microstructures 
and non-Newtonian behaviors of dense passive suspensions~\cite{seto2013discontinuous,Morris_2020}.
The main simulation methods are detailed in section~\ref{sec_method}. 
Section~\ref{sec_stripes}
presents the phenomenological results of the particle stripes, 
in terms of phase diagram and average stripe widths for various simulation conditions. 
Then, we give an insight into the mechanisms of forming the 
stripes (section~\ref{sec_forming_mechanism})
and predict the average stripe width from a theoretical scope 
(section~\ref{sec_theory}). 
At last, in section \ref{sec_rheology}, 
the influence of the striped microstructures
on the rheological property of the whole systems, 
involving shear viscosity, rotational viscosity, 
and the first normal stress coefficient, 
are studied.


\section{Simulation methods}
\label{sec_method}

\subsection{Particle dynamics}
We assume that a suspension of $N$ spherical particles, 
with a uniform radius $a$, 
are constrained in a monolayer ($x$-$y$ plane).
The forces and torques exerted on the particles 
($i=1,\dotsc,N$)
include those due to Stokes drag, $\Vec{F}_{\mathrm{S},i}$ and $\Vec{T}_{\mathrm{S},i}$,
those due to hydrodynamic (interparticle) interactions, 
$\Vec{F}_{\mathrm{H},ij}$ and $\Vec{T}_{\mathrm{H},ij}$, 
those due to frictional contact, $\Vec{F}_{\mathrm{C},ij}$ and $\Vec{T}_{\mathrm{C},ij}$, 
and those applied externally,
considered here to be a constant active torque that drives particle rotations, $\Vec{T}_{\mathrm{A}}$. 
The neglect of inertia and thermal agitations is justified 
for the case that the particle size is small and the applied shear stress is strong, 
i.e., the flow time scale is shorter than the Brownian time scale. 
Then, the force and torque balances on one particle are given by
\begin{gather}
  \Vec{F}_{\mathrm{S},i} 
  +
  \sum_{j \neq i} \left( 
  \Vec{F}_{\mathrm{H},ij} + \Vec{F}_{\mathrm{C},ij} \right)
  =   \Vec{0},
  \label{eq:force_balance} \\
  \Vec{T}_{\mathrm{S},i} 
  + 
  \sum_{j \neq i} \left( 
  \Vec{T}_{\mathrm{H},ij} + \Vec{T}_{\mathrm{C},ij} \right)
  +
  \Vec{T}_{\mathrm{A}}
  =
  \Vec{0}.
  \label{eq:torque_balance}
\end{gather}
The imposed simple shear flow can be expressed as
$\Vec{U}^{\infty}(\Vec{x}) = \Vec{\Omega}^{\infty}\times \Vec{x} +  \Tens{E}^{\infty} \cdot \Vec{x}$
in terms of the vorticity $2 \Vec{\Omega}^{\infty}$ and rate-of-strain tensor $\Tens{E}^{\infty}$,
where $\Vec{x}$ is the position vector. 
For a constant shear rate $\dot{\gamma}$, 
we have the nonzero elements $U^{\infty}_{x} = \dot{\gamma}y$, 
$\Omega^{\infty}_{z} = -\dot{\gamma}/2$, 
and $E^{\infty}_{xy} = E^{\infty}_{yx} = \dot{\gamma}/2$. 


The Stokes force and torque are given by
\begin{gather}
  \Vec{F}_{\mathrm{S},i}
  =
  -6 \pi \eta_0 a \left[ \Vec{U}_i - \Vec{U}^{\infty} (\Vec{x}_i) \right], 
  \label{eq:stokes_force} \\
  \Vec{T}_{\mathrm{S},i}
  =
  -8 \pi \eta_0 a^3 \left[ \Vec{\Omega}_i - \Vec{\Omega}^{\infty} (\Vec{x}_i) \right],
  \label{eq:stokes_torque}
\end{gather}
where $\eta_0$ is the solvent viscosity, 
and $\Vec{U}_i$ and $\Vec{\Omega}_i$ represent the velocity and angular velocity of particle $i$, respectively.
For the interparticle hydrodynamic interactions,
we assume they only come from lubrication effects. 
This is justified that for dense suspensions 
subjected to strong active torques and contact forces,
far-field or many-body hydrodynamic interactions play minor roles.
The lubrication force and torque, 
exerted by particles $j$ on $i$, 
then are given in the coupled form of 
\begin{equation}
  \begin{pmatrix}
    \Vec{F}_{\mathrm{H},ij}\\
    \Vec{T}_{\mathrm{H},ij}
  \end{pmatrix} 
  = 
-\Tens{R}_{\mathrm{L},ij}
\cdot
\begin{pmatrix}
\Vec{U}_j-
\Vec{U}^{\infty} (\Vec{x}_j) \\ 
\Vec{\Omega}_j-
\Vec{\Omega}^{\infty} (\Vec{x}_j)
 \end{pmatrix}
+
\Tens{R}'_{\mathrm{L},ij}
 : \Tens{E}^{\infty},
\end{equation}
where $\Tens{R}_{\mathrm{L},ij}$ and $\Tens{R}^{\prime}_{\mathrm{L},ij}$ are the resistance matrices 
for the hydrodynamic lubrication between particles $i$ and $j$~\cite{brady1988stokesian}.
In the current paper,
they can be simply described by the leading terms of the pairwise short-range lubrication interaction~\cite{ball1997}.
We also regularize the lubrication resistance 
to allow particle contacts~\cite{Wilson_2005}.


The interparticle contact force and torque are estimated 
using a simple spring-and-dashpot contact model~\cite{luding2008cohesive,mari2014shear}.
The force exerted by contacted particles $j$ on $i$ 
is composed of normal and tangential components,
given by
\begin{gather}
    \Vec{F}^{(\mathrm{n})}_{\mathrm{C},ij}
    =
    H(-h_{ij})
    (k_{\mathrm{n}} h_{ij} \Vec{n}_{ij} 
    + 
    \gamma_{\mathrm{n}} \Vec{U}^{(\mathrm{n})}_{ij}
    )
    ,
    \label{eq:contact_force_norm}\\
    \Vec{F}^{(\mathrm{t})}_{\mathrm{C},ij}
    =
 k_{\mathrm{t}} \Vec{\xi}_{ij},
    \label{eq:contact_force_tan}
\end{gather}
respectively.
Here $k_{\mathrm{n}}$ and $k_{\mathrm{t}}$ are the normal and tangential spring constants, respectively, 
$h_{ij}$ and $\Vec{n}_{ij}$ represent the surface separation 
and center-to-center unit vector (from $i$ to $j$) between the particles, respectively,
$\gamma_{\mathrm{n}}$ is the damping constant,
$\Vec{U}^{(\mathrm{n})}_{ij} \equiv \Vec{n}_{ij} \Vec{n}_{ij} \cdot \left( \Vec{U}_{j} - \Vec{U}_{i} \right)$ is the relative normal velocity, 
$H(x)$ is the Heaviside function,
and $\Vec{\xi}_{ij}$ denotes the tangential stretch vector.
Without contacting ($h_{ij} > 0$),
$\Vec{\xi}_{ij} = \Vec{0}$.
But after the particles contact at time $t_0$ ($h_{ij} \le 0$),
the tangential stretch vector evolves as $\Vec{\xi}_{ij}(t) =\int_{t_0}^{t} \Vec{U}^{\mathrm{(t)}}_{ij} dt$,
with the relative tangential velocity
$\Vec{U}^{\mathrm{(t)}}_{ij} \equiv (\Vec{I} - \Vec{n}_{ij} \Vec{n}_{ij}) \cdot 
[\Vec{U}_j - \Vec{U}_i - (a \Vec{\Omega}_i + a \Vec{\Omega}_j) \times \Vec{n}_{ij}]$.
For the contact torque,
it is obtained as 
\begin{equation}
    \Vec{T}_{\mathrm{C},ij}
    =
    a \Vec{n}_{ij} \times \Vec{F}^{(\mathrm{t})}_{\mathrm{C},ij}.
    \label{eq:contact_torque}
\end{equation}
Such a model exhibits a good description of contacts between gearlike particles~\cite{luding2008cohesive,mari2014shear}. 


The active torque $\Vec{T}_{\mathrm{A}}$ is 
perpendicular to the monolayer and uniformly acts on individual particles in the system.
For $\dot{\gamma} \neq 0$, 
we introduce a dimensionless measure
\begin{equation}
\tilde{T}_{\mathrm{A}} \equiv 
\frac{\Vec{T}_{\mathrm{A}} \cdot \Vec{e}_{z}
}{6 \pi \eta_{0} a^{3} \dot{\gamma} },
\label{eq:active_torque}
\end{equation}
where $\Vec{e}_{z}$ denotes the unit vector in the $z$ direction.
It is noted that $\tilde{T}_{\mathrm{A}}$ is a signed relative intensity of active torque 
to the shear rate of the imposed flow.
For convenience, we hereinafter call it the relative torque.
The positive and negative relative torques correspond to the particle rotations
in the opposite and same directions with respect to 
the vorticity, respectively.


\subsection{Simulation parameters and conditions}

Simulations are carried out for $N = 3000$ particles that are sheared with the Lees--Edwards periodic boundary conditions~\cite{Lees_1972}. 
We set $k_\mathrm{n}$ and $k_\mathrm{t}$ to sufficiently large values
that keep the maximum overlap and tangential displacement is less than $2 \%$ of the particle radius.
The relative torque and particle areal fraction are varied in the ranges of 
$-110 \le \tilde{T}_{\mathrm{A}} \le  110$ and $0.3 \le \phi \le 0.7$, respectively.


\subsection{Rheological characterization}
The stress tensor of the active suspension can be obtained as 
\begin{align}
    \Tens{\sigma} 
    &= 
        2 \eta_0 \Tens{E}^{\infty}
    - \frac{1}{V} \sum_{i > j}
    \Vec{r}_{ij} 
    \left(
    \Vec{F}_{\mathrm{H},ij}
    +
    \Vec{F}_{\mathrm{C},ij}
    \right) 
    \notag \\
    & = 
    2 \eta_0 \Tens{E}^{\infty}
    - 
    \frac{1}{V} \mathrm{sym}
    \sum_{i > j}
        \Vec{r}_{ij} 
        \left(
        \Vec{F}_{\mathrm{H},ij}
        +
        \Vec{F}_{\mathrm{C},ij}
        \right) 
        \notag \\
    &\quad  + \frac{N}{V} \Tens{\epsilon} \cdot 
    \left\{ \Vec{T}_{\mathrm{A}} - 8 \pi \eta_0 a^3 ( \langle \Vec{\Omega} \rangle 
    - 
    \Vec{\Omega}^{\infty} ) 
    \right\},
    \label{eq:stress_tensor}
\end{align}
where the symbol of $\mathrm{sym}$ represents the symmetric part of the tensor, 
$V$ the total volume of the suspension, 
$\Vec{r}_{ij}$ ($\equiv \Vec{x}_{j} - \Vec{x}_{i}$) the center-to-center vector between particles $i$ and $j$,
$\Tens{\epsilon}$ the rank-3 Levi-Civita tensor,
and $\langle \Vec{\Omega} \rangle$ the average angular velocity of the particles.
Note that
the stress tensor
$ \Tens{\sigma}$
is nonsymmetric
when $\Vec{T}_{\mathrm{A}} \neq 8 \pi \eta_0 a^3 ( \langle \Vec{\Omega} \rangle - \Vec{\Omega}^{\infty} )$.
Then, the shear viscosity and 
the first normal stress coefficient can be calculated by 
\begin{gather}
    \eta 
    \equiv 
    \frac{(\mathrm{sym}\,\Tens{\sigma})_{xy}}{\dot{\gamma}},
    \label{eq:shearviscosity}\\
    \Psi_1
    \equiv
    \frac{N_{1}}{|\dot{\gamma}|},
    \label{eq:normalstressdifference}
\end{gather}
respectively,
where $N_1 \equiv \sigma_{xx} - \sigma_{yy}$ is the first normal stress difference. 
If the microstructure 
of non-Brownian hard sphere suspensions
does not depend
on the shear rate,
the normal stresses are proportional
to the shear stress~\cite{seto2018normal}.
Thus, $\Psi_1$ is defined similarly to $\eta$~\cite{dbouk2013normal} instead of the alternative way,
i.e., $N_1/\dot{\gamma}^2$~\cite{larson1999structure}.


The rotational viscosity, 
which describes the damped transfer of angular momentum from particles to the surrounding solvent,
can be obtained from the asymmetric component 
of the stress tensor\,\cite{dahler1963theory,brenner1970rheology}, as 
\begin{align}
    \eta_{\mathrm{rot}} 
    \equiv 
    \frac{(\mathrm{asym}\,\Tens{\sigma})_{xy}}{ 2 ( \langle \Omega_z \rangle - \Omega_z^{\infty} ) } 
    =
    2 \phi \eta_0
    \left(
    \frac{3 \tilde{T}_{\mathrm{A}}} 
    {2 ( \langle \tilde{\Omega}_{z} \rangle 
    - 1  )} - 1
    \right),
    \label{eq:rotationalviscosity}
\end{align}
where $\mathrm{asym}$ represents the asymmetric part of the tensor
and 
$ \langle \tilde{\Omega}_{z} \rangle \equiv \langle \Omega_{z} \rangle / |\Omega^{\infty}_{z}| $.
It is noted that the rotational viscosity is indefinable
when the active torque becomes zero 
and, thus, $\langle \tilde{\Omega}_{z} \rangle  = 1$.
The derivations of the stress tensor and rotational viscosity can be found in the Appendix.

In addition, we introduce the so-called odd viscosity $\lambda_{\mathrm{odd}}$
within the steady shear characterization,
which essentially is relevant to what we defined above.
The odd viscosity 
is usually addressed as a peculiar rheological nature 
of chiral viscous fluids\,\cite{soni2019odd,hargus2020time,han2021fluctuating}.
In terms of usual definition, 
it is one element of tensorial viscosity $\eta_{ijkl}$
(in a generalized viscous model, 
$\sigma_{ij}= \eta_{ijkl} v_{kl}$) with the base $\epsilon_{ik}\delta_{jl}+\epsilon_{jl}\delta_{ik}$,
where $v_{ij}$
and $\delta_{ij}$
are the velocity gradient tensor and Kronecker delta,
respectively\,\cite{epstein2020time}.
To identify the odd viscosity from rheological measurements,
we need to evaluate $\Psi_1$
under positive and negative shear rates
and take the odd part of them,
\begin{equation}
    \lambda_{\mathrm{odd}} 
    \equiv \frac{1}{4}
    \left\{
    \Psi_1(\dot{\gamma}) - \Psi_1(-\dot{\gamma})
    \right\}.
\end{equation}
The remaining even part is then given by 
\begin{equation}
    \lambda_{\mathrm{even}} 
    \equiv \frac{1}{4}
    \left\{
    \Psi_1(\dot{\gamma}) + \Psi_1(-\dot{\gamma})
    \right\}.
\end{equation}
Both $ \lambda_{\mathrm{odd}} $ and $ \lambda_{\mathrm{even}} $ only measure normal stress anisotropy \citep{seto2018normal},
but do not associate with energy dissipation.
We note that $ \lambda_{\mathrm{even}} $ is equivalent to $\Psi_1$ for non-chiral fluids 
and not necessarily zero for chiral fluids in general.
However, for purely viscous fluids satisfying objectivity, 
it should be zero~\cite{Phan-Thien_2017}.


\begin{figure}[tb]
  \includegraphics[width=0.47\textwidth]{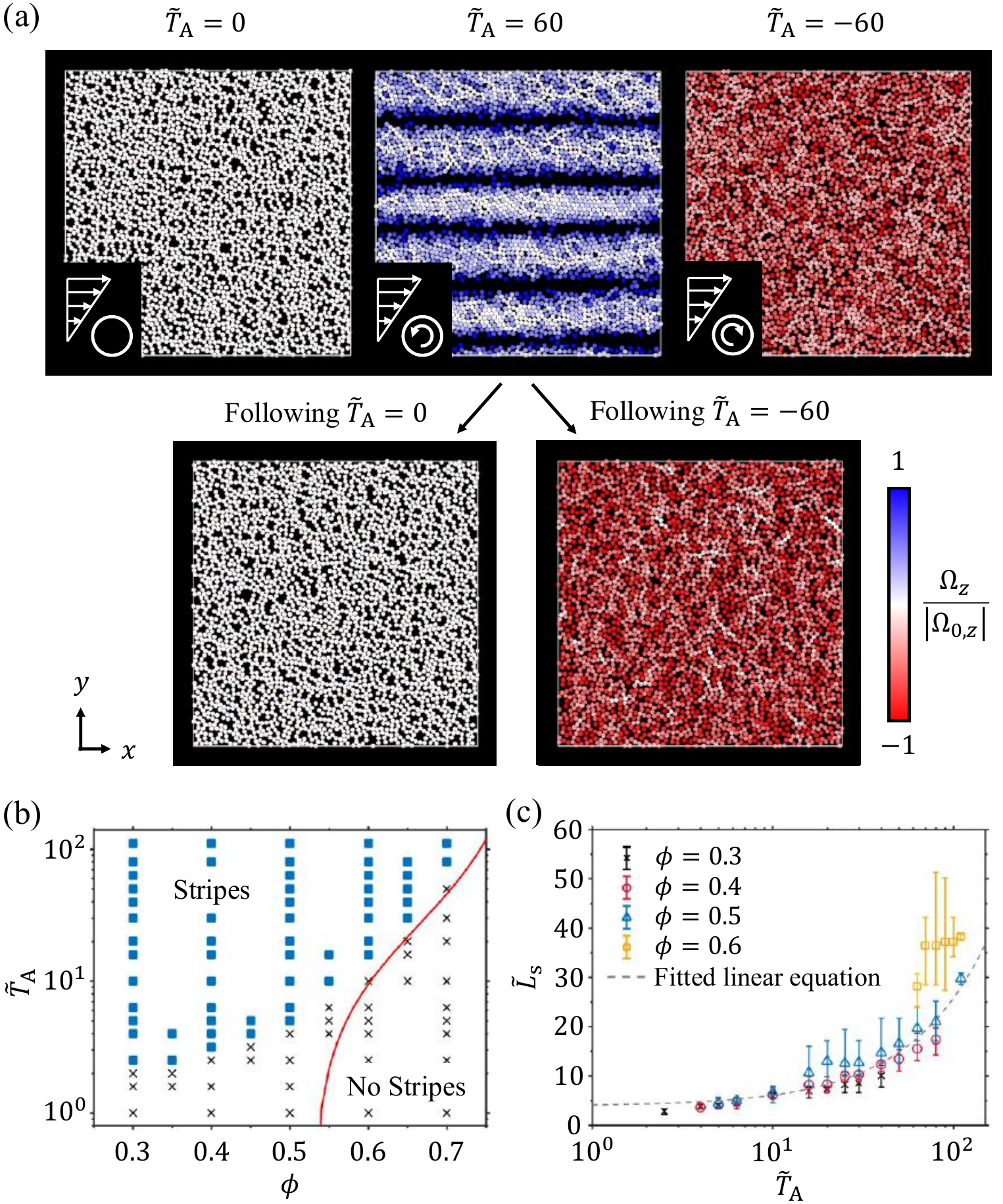}
   \caption{\label{fig:characterization}%
   (a) Representative snapshots of configuration of particles subjected to zero, positive, and negative relative torques.
   The particles subjected to the positive torque also experience following zero and negative torques. 
   Here the particle areal fraction $\phi = 0.6$.
   The color of the particle denotes the relative angular velocity of the particle 
   to the angular velocity of a single particle under $\tilde{T}_{\mathrm{A}}$. 
   The insets denote the direction of self-rotation of the particles (represented by the arrow within the circle)
   with respect to the imposed simple shear flow. 
   The circle without an arrow represents the passive particle. 
   (b) Nonequilibrium phase diagram for positive relative torques. 
   The red line represents the phase boundary,
   which is predicted based on the fitted linear equation from (c) and the assumptions that
   the particle areal fraction within the stripes and average interstripe separation are $0.8$ and $2a$, respectively. 
   (c) Average width of particle stripes as a function of intensity of relative torque for various particle areal fractions.
   The simulation data is exhibited with error bars.
   The dashed line is a linear equation that is fitted from the data for $0.3 \le \phi \le 0.5$.}
\end{figure}


\section{Results}

\subsection{Stripes of self-rotating particles}

\label{sec_stripes}
\Figrefp{fig:characterization}{a} shows the fully developed particle configurations
for $\phi = 0.6$ and different relative torques.
In the figure, the color of the particle visualizes the relative angular velocity of the particle to the angular velocity of a single particle under $\tilde{T}_{\mathrm{A}}$ (i.e., $\Omega_{0,z}$).
The insets illustrate the direction of self-rotation of the particles with respect to the simple shear flow.
We observe that the particles driven by the positive relative torque ($\tilde{T}_{\mathrm{A}} = 60$) form prominent stripy aggregates, 
with faster-rotating particles on the edges and slower-rotating particles inside the stripes. 
A following sudden suppression or reversal of direction of the torque is able to rerandomize the particles. 


Such an intriguing phenomenon, however, is not seen for the case of zero or negative relative torque,
where the particles are randomly distributed throughout. 
Additionally, we examine the configuration of self-rotating particles in planar extensional flows.
The result, as shown in 
\figref{fig_A1} of \Appendref{app:figures},
suggests that the pure shear flow cannot induce any particle aggregates.
Since the simple shear flow is a superposition of pure shear and vortical flows 
and the latter one is difficult to be modeled in simulations, 
we then infer that the vortical component should be necessary for constructing the particle stripes. 
More particle configurations in simple shear flows and for various values of $\phi$ and $\tilde{T}_{\mathrm{A}}$ 
can be found in \figrefs{fig_A2}{fig_A4} of \Appendref{app:figures}.


By taking an overview of particle configurations under positive relative torques, 
we obtain the nonequilibrium phase diagram as shown in \figrefp{fig:characterization}{b}. 
In the figure, the black crosses denote the configurations featured by randomly distributed particles,
whereas the blue squares denote the configurations consisting of particle stripes. 
The result shows that the particle stripes only appear when the value of $\tilde{T}_{\mathrm{A}}$ is greater than certain $\phi$-dependent thresholds. 
In addition, further increasing the value of $\tilde{T}_{\mathrm{A}}$ leads to 
a transfer of particle structure from crystalline to uncrystalline states 
(see \figref{fig_A2} of \Appendref{app:figures}).


Focusing on the configurations consisting of fully developed particle stripes,
we study the average stripe width $\tilde{L}_{\mathrm{s}}$ (scaled by the particle radius) 
as a function of intensity of the relative torque. 
As seen in \figrefp{fig:characterization}{c}, when $0.3 \le \phi \le 0.5$, 
the dependence of $\tilde{L}_{\mathrm{s}}$ on $\tilde{T}_{\mathrm{A}}$ 
follows a linear equation of $\tilde{L}_{\mathrm{s}} \approx 0.22 \tilde{T}_{\mathrm{A}} + 3.92 $, 
which is independent of $\phi$. 
Increasing the value of $\phi$ only results in the decrease of stripe number and interstripe gap
(see \figref{fig_A6} of \Appendref{app:figures}). 
When the gap is smaller than the particle diameter, 
the preformed stripes combine in part or throughout the system 
(see \figref{fig_A7} of \Appendref{app:figures}).
This explains for the case of $\phi = 0.6$,
where $\tilde{L}_{\mathrm{s}}$ undergoes a steplike increase with $\tilde{T}_{\mathrm{A}}$ 
and its value is larger than (even twice of) the prediction of the linear equation.


Moreover, based on the fitted linear equation and assumption that the packing fraction of in-stripe particles is consistently 0.8, 
we theoretically predict the phase boundary in \figrefp{fig:characterization}{b} (see the red solid line).
Such a prediction shows a good agreement with the simulation results for $\phi \ge 0.6$,
but deviates when $\phi \le 0.55$.
This can be explained by the decreased packing fraction of the in-stripe particles for the latter case. 
Please see \Appendref{app:phase_boundary} for the theoretical derivation 
and the study of the phase boundary for various packing fractions.


\begin{figure}[tb]
\includegraphics[width=0.47\textwidth]{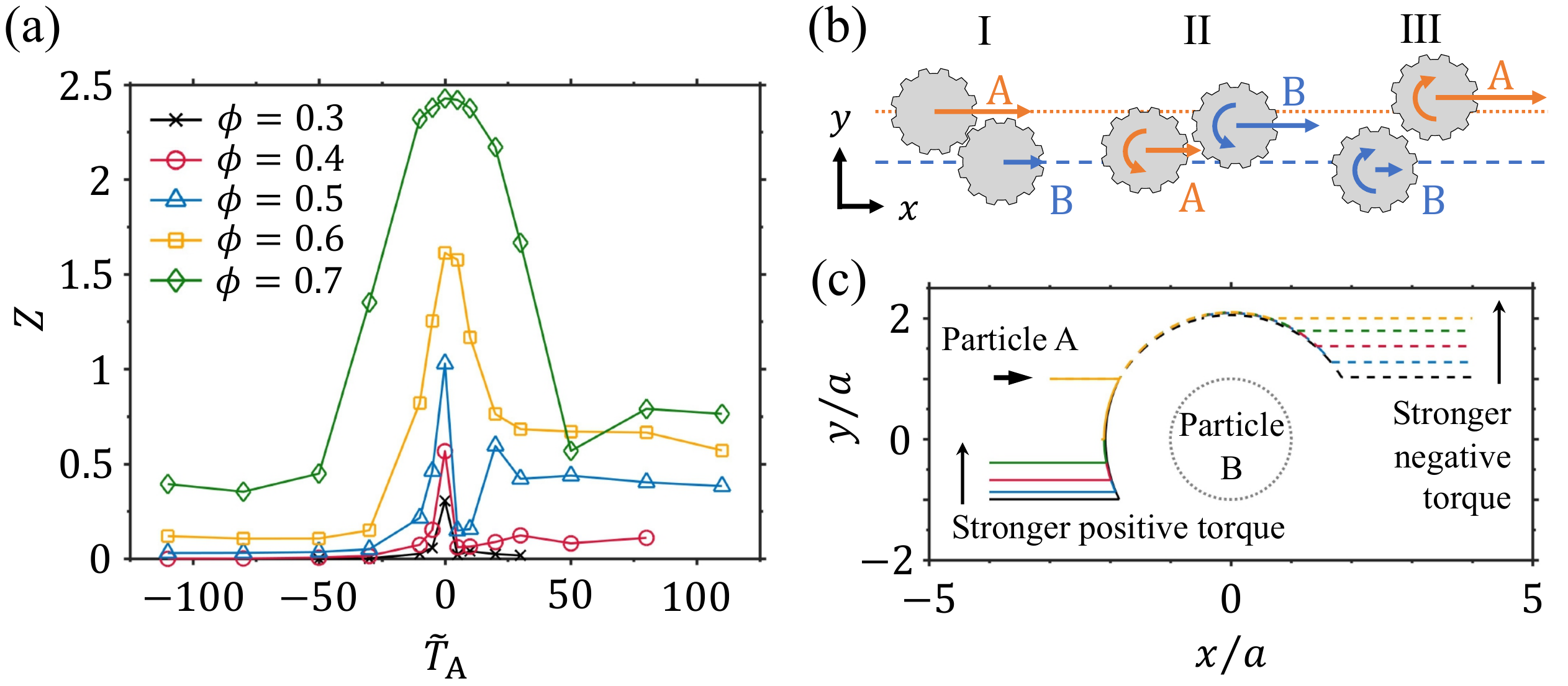}
\caption{\label{fig:rotational_rearrangement}%
(a) Average contact number $Z$ as a function of relative torque for various particle areal fractions.
(b) Illustration of pairwise interaction. 
Sketch I shows the initial particle configuration, 
whereas II and III show the particle configurations after the rotation-induced rearrangement 
for the cases of positive and negative relative torques, respectively. 
The orange dotted and blue dashed lines indicate the initial $y$ positions of particles A and B, respectively.
(c) Relative trajectories of particle A with respect to particle B for various relative torques.}
\end{figure}

\begin{figure*}
\includegraphics[width=0.95\textwidth]{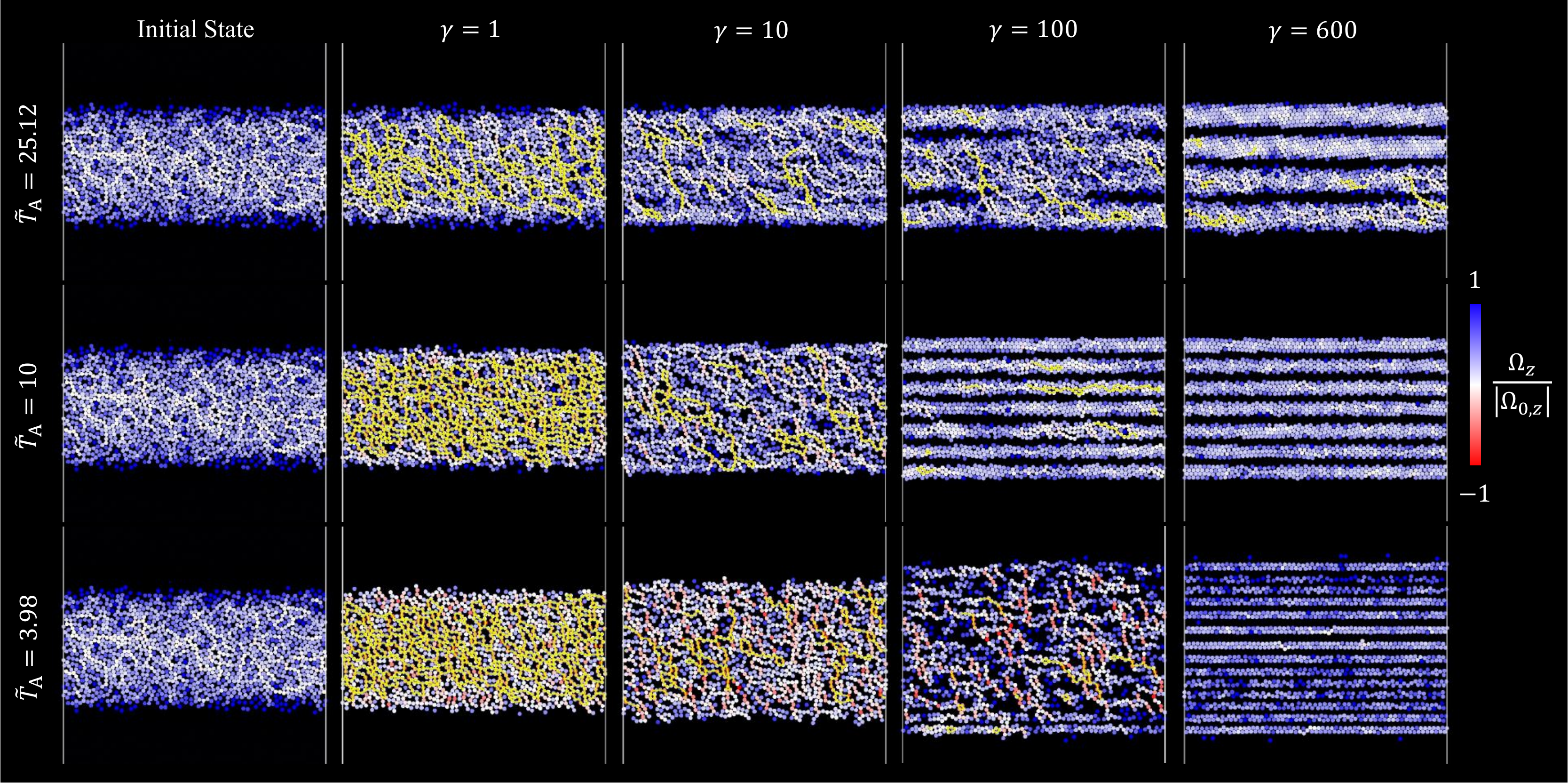}
\caption{\label{fig:expansion}%
Representative snapshots of particle configurations showing the relaxation of a thick particle stripe.
The thick particle stripe is initially stable for $\tilde{T}_{\mathrm{A}} = -140$ 
and suddenly subjected to the weakened torques $\tilde{T}_{\mathrm{A}} = -25.12$, $-10$, and $-3.98$. 
Here $\gamma$ represents the shear strain,
which is proportional to the simulation time.
The color of the particle denotes the relative angular velocity of the particle 
to the angular velocity of a single particle under $\tilde{T}_{\mathrm{A}}$. 
The yellow bonds between the particles indicate the particle contacts (or contact chains).}
\end{figure*}


\subsection{Mechanisms of forming particle stripes}
\label{sec_forming_mechanism}
In this section, we study three typical collective behaviors of the self-rotating particles in simple shear flows,
i.e., shear-induced diffusion, rotation-induced rearrangement, and edge flows.
Based on them, 
we explain how the particle stripes are formed and the stripe width is determined.


It is known that shear flows can cause interparticle contacts and 
even formation of chainlike clusters along the compressive axis
for dense and passive particles with rough surfaces\,\cite{seto2013discontinuous,mari2014shear}.
Within such a process, 
the contact forces break 
the reversible nature of the Stokes flows\,\cite{Wilson_2000}.
The clusters rotate under shear,
resulting in the spread and then more uniform distribution of the particles 
along the direction of the velocity gradient. 
Since the structures of the clusters are irregular, 
such an effectively ``random'' drift motion of particles is also called shear-induced diffusion,
which has been extensively studied by prior works\,\cite{leighton1987measurement,sierou2004shear}.


However, the chainlike clusters are not commonly seen 
for the dense suspensions of frictional self-rotating particles 
(except for the situations of very large areal fractions and small values of $|\tilde{T}_{\mathrm{A}}|$, 
as shown in 
\figrefs{fig_A3}{fig_A5} of \Appendref{app:figures}). 
By examining the average number of particle contact $Z$ [see \figrefp{fig:rotational_rearrangement}{a}], 
we observe that 
the self-rotating particles are always less in contact than the passive particles.
Additionally, increasing the value of $|\tilde{T}_{\mathrm{A}}|$ 
leads the contact number to decrease before particle stripes appear, 
but to be changeless when the stripes are emergent. 
Compared with the particles driven by negative torques, 
those within stripes are more in contact. 


To explain the effect of active torque on the particle contacts, 
the motion of two interacting particles is investigated, 
as shown in \figrefp{fig:rotational_rearrangement}{b}.
Let us consider two corotating particles, A and B, 
are placed in a simple shear flow,
with an initial $y$-directional separation that is equal to the particle radius.
Since along the $x$ direction particle A translates faster than particle B,
they approach until the interparticle hydrodynamic lubrication and even frictional contact force become dominant.
Such a moment is illustrated by sketch I in \figrefp{fig:rotational_rearrangement}{b}, 
where the orange dotted and blue dashed lines indicate 
the initial $y$ positions of particles A and B, respectively. 
Then, due to the tangential contribution of the interacting force, 
the particles undergo a rotation-induced rearrangement,
which relieves the particle contacts and, more importantly,
induces the $y$-directional effective attraction (see sketch II) and repulsion (see sketch III) 
for the particles driven by the positive and negative relative torques, respectively.
These phenomena are numerically confirmed by 
the relative trajectories of particles A to B in \figrefp{fig:rotational_rearrangement}{c}.
Moreover, considering the shear-induced diffusion can generate a $y$-directional effective repulsion,
it could be competitive and synergistic with the effective attraction and repulsion by the rotation-induced rearrangement, respectively. 
Then the two-dimensional particle dynamics simply becomes one-dimensional ($y$-directional), 
and the motivation of the particles driven by positive torques to align along the $x$ direction is reasoned. 
However, in order to explore the role of active torque, 
\figrefp{fig:rotational_rearrangement}{b} only exhibits the case 
where the rotation-induced rearrangement is dominant. 


In \figrefp{fig:rotational_rearrangement}{c}, 
it is also observed that effective interactions are enhanced for large relative torques.
Such a result implies the equivalent role of the relative torque to describe the competition
between the rotation-induced rearrangement and shear-induced diffusion.
To give insight into such a competition,
we carry out simulations for the particles driven by the positive relative torques, 
where a thick particle stripe is initially stable under $\tilde{T}_{\mathrm{A}} = 140$ 
and suddenly subjected to weakened torques,
as shown in \figref{fig:expansion}.
In the figure, $\gamma$ denotes the shear strain
and the interparticle contacts (or contact chains) are visualized by the highlighted yellow bonds.
We observe that immediately after weakening the relative torque ($\gamma = 1$), 
dense contact chains appear within the stripe,
although the stripe width hardly changes.
This is explained by the weakened rotation-induced rearrangement,
which cannot efficiently relieve the particle contacts as before 
and only induces weak effective attractions.
As a result, the shear-induced diffusion impels the particles to expand in the $y$ direction, 
where more free space is available, 
and then organize into multiple narrow stripes based on the decreased local density.
Compared with the case of $\tilde{T}_{\mathrm{A}} = 25.12$, 
the weaker positive relative torques lead to stronger shear-induced diffusion, 
which gives rise to more crowded contact chains at the beginning and more and thinner particle stripes eventually. 
According to such results,
we highlight the importance of competition between the shear-induced diffusion and rotation-induced rearrangement
on determining the formation and average width of particle stripes.

\begin{figure}[tb]
  \includegraphics[width=0.47\textwidth]{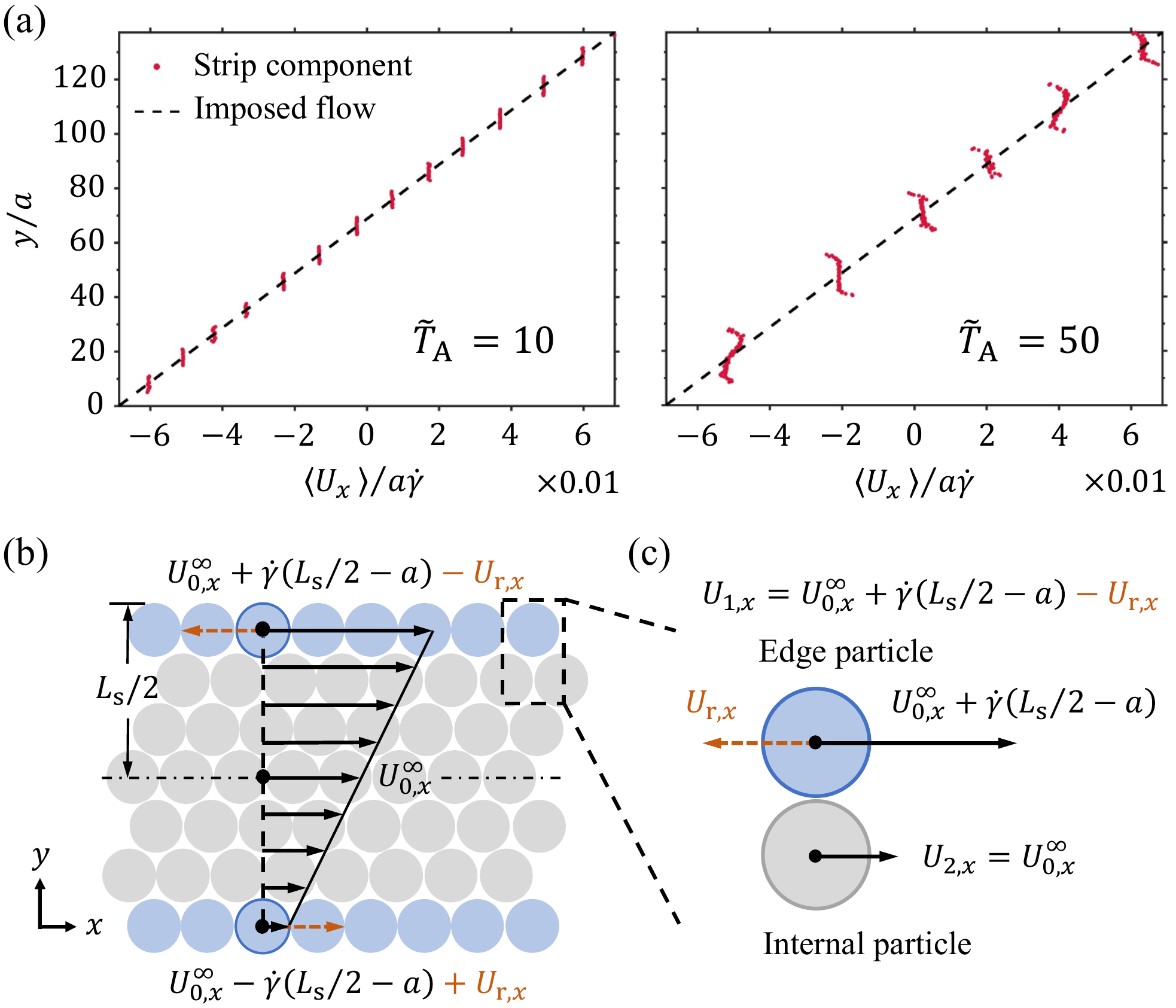}
  \caption{\label{fig:stripe_analysis}%
  (a) Profiles of average $x$ component velocity of particles for $\phi = 0.5$ and various values of $\tilde{T}_{\mathrm{A}}$.
  Schematic illustrations of (b) particle stripe and (c) interaction model of internal and edge particles
  (on the top edge for instance). 
  The internal particles, denoted in grey, 
  are assumed to be a rigid body, 
  translating with the flow velocity at the middle height of the stripe $U^{\infty}_{0,x}$.
  However, the edge particles, denoted in blue, 
  experience two $x$ contributions of velocity: 
  that is carried by the background flow $U^{\infty}_{0,x} + \dot{\gamma} (L_{\mathrm{s}}/2 - a)$,
  and that is induced by the self-rotation $U_{\mathrm{r}}$.}
\end{figure}


A further insight is given into the effect of interaction forces on the particle configuration.
In \figref{fig_A8} of \Appendref{app:figures}, 
one can observe that 
application of only frictional contact (without hydrodynamic lubrication) contributes to the formation of crystallized stripes, 
whereas with only hydrodynamic lubrication (without frictional contact) the particles construct uncrystalline stripes.
Combining with the observations in \figref{fig_A2} of \Appendref{app:figures} 
(that weak and strong relative toques lead to crystalline and uncrystralline stripes, respectively),
we point out that under the weak and strong relative toques, 
the interparticle interaction is dominated by the hydrodynamic lubrication and frictional contact, respectively.

\Figrefp{fig:stripe_analysis}{a} shows two representative $x$-directional velocity profiles of fully developed particle stripes for $\phi = 0.5$, 
where the black dashed line denotes the velocity of the imposed flow.
The $x$-directional velocity $\langle U_{x} \rangle$ is estimated by taking the average of particle velocities within differential $y$ scales.
For $\tilde{T}_{\mathrm{A}} = 10$,
we see that each stripe translates like a rigid body, 
with the flow velocity at the middle height of the stripe. 
However, for $\tilde{T}_{\mathrm{A}} = 50$ the velocity profiles deform,
with the velocity at the stripe boundary antiparallel to the local flow velocity.


To explain the above phenomena, 
we divide the in-stripe particles into two groups: 
those on the edge of the stripe and those stay internally, 
as seen in \figrefp{fig:stripe_analysis}{b}.
For simplification,
a single layer of the edge particles is assumed. 
Because the internal particles are locally confined (or jammed), 
they have reduced rotational mobility and translate like a rigid body, 
with the flow velocity at the middle height of the stripe $U^{\infty}_{0,x}$.
By interacting with such a rigid body,
the edge particle then generate a rotation-induced velocity $U_{\mathrm{r},x}$,
which is $\tilde{T}_{\mathrm{A}}$-dependent and always antiparallel to the local flow velocity.
Additional to the local flow velocity, 
the resultant velocity of the top and bottom edge particles are given by $U^{\infty}_{0,x} + \dot{\gamma} (L_{\mathrm{s}}/2 - a) - U_{\mathrm{r},x}$ and $U^{\infty}_{0,x} - \dot{\gamma} (L_{\mathrm{s}}/2 - a) + U_{\mathrm{r},x}$, respectively.
Here $\dot{\gamma} (L_{\mathrm{s}}/2 - a)$ represents the difference of the flow velocity from the edge particle location to the middle height of the stripe.
When $U_{\mathrm{r},x} = \dot{\gamma} (L_{\mathrm{s}}/2 - a)$,
the entire stripe undergoes a rigidlike translation,
as the velocity profile shown in \figrefp{fig:stripe_analysis}{a} for $\tilde{T}_{\mathrm{A}} = 10$.
However, $U_{\mathrm{r},x} > \dot{\gamma} (L_{\mathrm{s}}/2 - a)$ leads to significant edge flows of the particles,
as that in the deformed velocity profile for $\tilde{T}_{\mathrm{A}} = 50$. 
In addition, we observe that the stripes are stable only when $U_{\mathrm{r},x} \ge \dot{\gamma} (\tilde{L}_{\mathrm{s}}/2 - a)$.
Otherwise, they deform in the same way as shown in \figref{fig:expansion}.
The fully developed $x$-directional velocity profiles for 
additional $\phi$ and $\tilde{T}_{\mathrm{A}}$ can be found in \figref{fig_A9} of \Appendref{app:figures}.


\subsection{Dependence of stripe width on relative torque}
\label{sec_theory}
We confirm the linear relation between the average width of particle stripes and relative torque,
by conducting a theoretical study based on the particle dynamics.
As shown in \figrefp{fig:stripe_analysis}{b}, 
we consider that a particle stripe with width $L_{\mathrm{s}}$ is fully developed in a simple shear flow with shear rate $\dot{\gamma}$.
The internal particles translate at the uniform velocity 
that is equal to the flow velocity at the middle height of the stripe $U^{\infty}_{0,x}$,
whereas the edge particles have an additional rotationally induced velocity $U_{\mathrm{r},x}$ against the local flows.


Since the interactive effects from the neighbor particles in the $x$ direction are counteracted,
the model can be simplified to the pairwise interaction 
between one edge (noted by subscript $1$) and one internal (noted by subscript $2$) particles,
as seen in \figrefp{fig:stripe_analysis}{c}. 
In addition, an overview of configurations of fully developed particle stripes suggests that 
the edge and internal particles are hardly in contact. 
Thus, for the edge particle, Eqs.~\eqref{eq:force_balance} and \eqref{eq:torque_balance} reduce to
\begin{equation}
  \Vec{F}_{\mathrm{S},1}
  +
  \Vec{F}_{\mathrm{H},12}
  =   \Vec{0}
\quad\text{and}\quad
  \Vec{T}_{\mathrm{S},1}
  +
  \Vec{T}_{\mathrm{H},12}
  +
  \Vec{T}_{\mathrm{A}}
  =
  \Vec{0}.
  \label{eq:force_torque_balance_reduced}
\end{equation}


We are only interested in the force balance in the $x$ direction and torque balance in the $z$ direction
\begin{gather}
    \begin{aligned}
        -6 \pi \eta_0 a  \left( U_{1,x} - U^{\infty}_{1,x} \right) 
        +
        R^{F U}_{\mathrm{L}} \left( U_{1,x} - U_{2,x} \right) \\
        \phantom{A} + 
        R^{F \Omega}_{\mathrm{L}} \left( \Omega_{1,z} - \Omega_{2,z} \right)
        = 0,
    \end{aligned}
    \label{eq:force_balance_derive} \\
    \begin{aligned}
        -8 \pi \eta_0 a^3  \left( \Omega_{1,z} - \Omega^{\infty}_{1,z} \right)
        +
        R^{T U}_{\mathrm{L}} \left( U_{1,x} - U_{2,x} \right) \\
        \phantom{A} + 
        R^{T \Omega}_{\mathrm{L}} \left( \Omega_{1,z} - \Omega_{2,z} \right)
        +
        T_{\mathrm{A},z}
        = 0,
    \end{aligned}
    \label{eq:torque_balance_derive}
\end{gather}
where $U^{\infty}_{1,x}$ and $\Omega^{\infty}_{1,z}$ are the $x$  and $z$ components of the velocity and angular velocity of the imposed flow at the position of the edge particle, respectively.
In Eqs.~\eqref{eq:force_balance_derive} and~\eqref{eq:torque_balance_derive}, 
$R^{F U}_{\mathrm{L}}$, $R^{F \Omega}_{\mathrm{L}}$, $R^{T U}_{\mathrm{L}}$, and $R^{T \Omega}_{\mathrm{L}}$ 
are the resistance coefficients coupling force and torque to velocity and angular velocity through the hydrodynamic lubrication. 
They are functions of interparticle separation but not of $\tilde{T}_{\mathrm{A}}$ or $\tilde{L}_{\mathrm{s}}$.
For the internal particle, 
we set $\Omega_{2,z} = 0$ and $U_{2,x} = U^{\infty}_{0,x}$.
The definition of $\Delta U \equiv U_{1,x} - U_{2,x}$ also leads to
\begin{equation}
    \begin{aligned}
        U_{1,x} - U^{\infty}_{1,x} 
        & =
        U_{1,x} - \{ U^{\infty}_{0,x} + \dot{\gamma} ( L_{\mathrm{s}} / 2 - a ) \} \\
        & =
        \Delta U -\dot{\gamma} ( L_{\mathrm{s}} / 2 - a ).
    \end{aligned}
    \label{eq:velocity_difference}
\end{equation}
By substituting Eqs.~\eqref{eq:torque_balance_derive} and\,\eqref{eq:velocity_difference} into\,\eqref{eq:force_balance_derive},
we then eliminate $\Omega_{1,z}$ and obtain 
\begin{equation}
    \tilde{L}_{\mathrm{s}}
    =
    \alpha \tilde{T}_{\mathrm{A}}
    +
    2 ( 1 - \alpha/3 )
    +
    \beta \Delta \tilde{U},
    \label{eq:linear_first_principle}
\end{equation}
where $\Delta \tilde{U} = \Delta U / a \dot{\gamma}$,
\begin{gather}
    \alpha
    =
    \frac{ 2 a R^{F \Omega}_{\mathrm{L}} } { R^{T \Omega}_{\mathrm{L}} - 8 \pi \eta_0 a^3 },
    \label{eq:linear_first_principle_alpha} \\
    \beta
    =
    2
    +
    \frac{ \alpha R^{TU}_{\mathrm{L}} - 2 a R^{FU}_{\mathrm{L}} } { 6 \pi \eta_0 a^2 }.
    \label{eq:linear_first_principle_beta}
\end{gather}
It should be noted that unlike $\alpha$ and $\beta$, 
$\Delta \tilde{U}$ is a function of $\tilde{L}_{\mathrm{s}}$ and $\tilde{T}_{\mathrm{A}}$.


When the entire stripe is rigidlike [as the case of $\tilde{T}_{\mathrm{A}} = -10$ in \figrefp{fig:stripe_analysis}{a}], 
the particle velocities satisfy $U_{1,x} = U_{2,x}$, i.e., $\Delta \tilde{U} = 0$. 
Thus, 
Eq.~\eqref{eq:linear_first_principle} becomes
\begin{equation}
    \tilde{L}_{\mathrm{s}}
    =
    \alpha \tilde{T}_{\mathrm{A}}
    +
    2 ( 1 - \alpha/3 ),
    \label{eq:linear_first_principle_rigid_body}
\end{equation}
which shows the linear relation between  $\tilde{L}_{\mathrm{s}}$ and $\tilde{T}_{\mathrm{A}}$. 
Since the above theoretical analysis is based on the fully developed particle stripes,
Eqs.~\eqref{eq:linear_first_principle} and\,\eqref{eq:linear_first_principle_rigid_body} break down 
for zero and positive relative torques as well as weak negative relative torques that cannot cause particle stripes. 


\begin{figure*}[ht]
\includegraphics[width=0.90\textwidth]{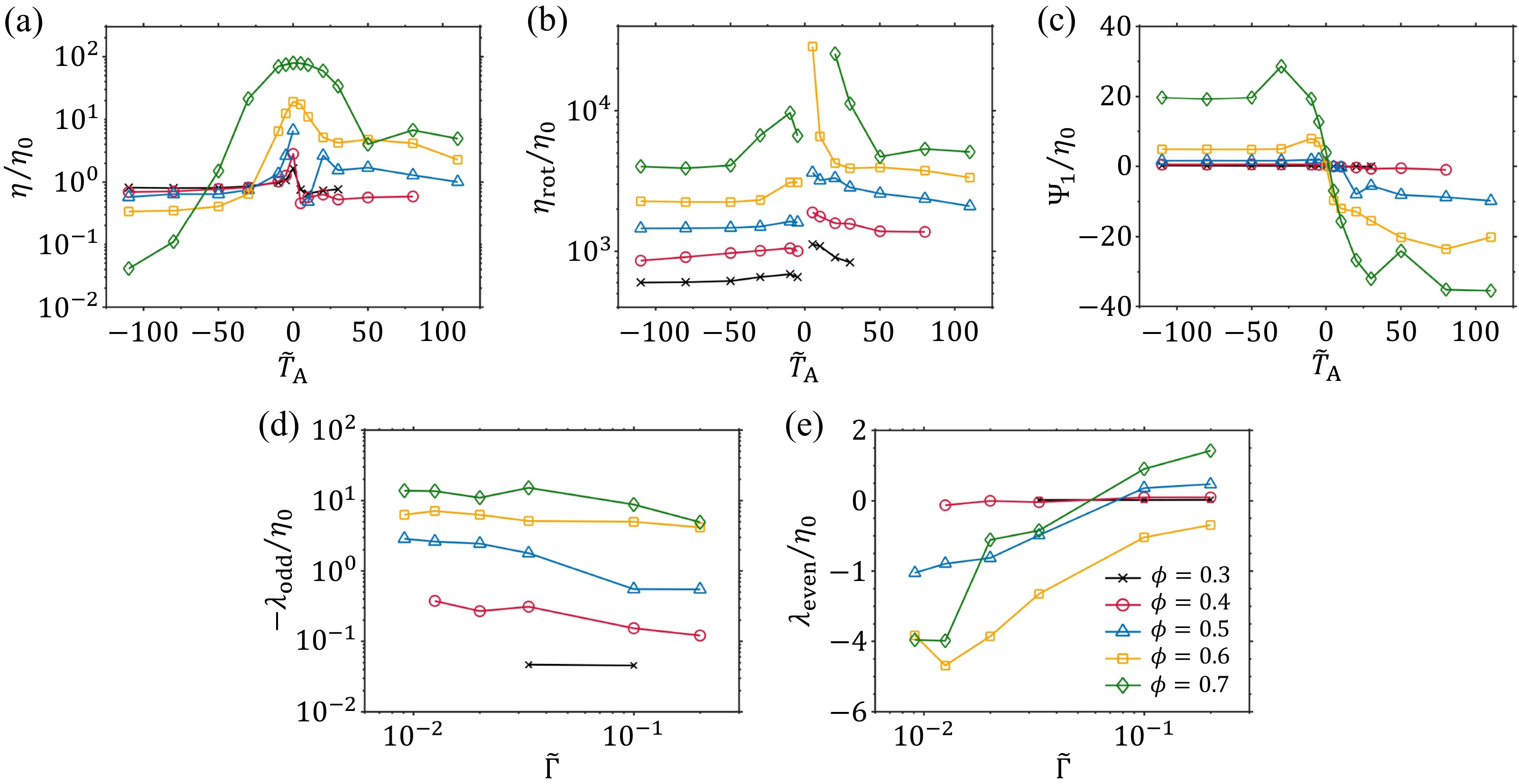}
\caption{\label{fig:rheology} (a) Shear viscosity $\eta$, 
(b) rotational viscosity $\eta_{\mathrm{rot}}$, 
and
(c) the first normal stress coefficient $\Psi_1$
of the rotor suspension 
are plotted 
as a function of relative torque $\tilde{T}_{\mathrm{A}}$ 
for various particle areal fractions $\phi$
with $\dot{\gamma}>0$.
To quantify the rheological oddness,
(d) the odd viscosity $\lambda_{\mathrm{odd}}$ and (e) even part $\lambda_{\mathrm{even}}$ of the first normal stress coefficient 
are plotted
as a function of
the dimensionless shear rate 
$\tilde{\Gamma} \equiv \dot{\gamma}/(|\Vec{T}_{\mathrm{A}} \cdot \Vec{e}_{z}| / 6\pi \eta_0 a^3)$
for $\Vec{T}_{\mathrm{A}} \cdot \Vec{e}_z > 0$.
Here all the rheological measures are nondimensionalized by the solvent viscosity $\eta_0$.}
\end{figure*}


\subsection{Effect of particle stripes on rheology}
 \label{sec_rheology}
The influence of 
the shear-induced microstructures
on the (macroscopic) suspension rheology is studied in terms of 
shear viscosity $\eta $, 
rotational viscosity $\eta_{\mathrm{rot}}$, 
and the first normal stress coefficient $\Psi_{1}$.
Each data point in \figref{fig:rheology} corresponds to a single run of different simulation conditions, 
and is averaged over certain simulation times after the particle configurations are fully developed.
We discuss the results for both positive relative torques that lead the striped microstructure 
and zero and negative relative torques for comparison.
As observed in \figref{fig:rheology}, 
increasing the value of $|\tilde{T}_{\mathrm{A}}|$ 
leads both shear and rotational viscosities to decrease,
and the decrease for the negative torques is larger than that for the positive torques. 


For a fixed active torque in any directions, 
\figrefp{fig:rheology}{a} 
indicates the shear thickening of rate dependence.
Although the similar phenomenon of shear viscosity is also seen for dilute suspensions of Quincke rotors 
that rotate in the same direction of imposed vorticity~\cite{lobry1999viscosity},
we note their origins are different. 
In our system, the phenomenon is explained that 
for large values of $|\tilde{T}_{\mathrm{A}}|$ the enhanced positional rearrangements 
can efficiently release particle contacts,
and the particles driven by the positive torques may still contact within the stripes.
The existence of such particle contacts, or resultant shear-induced microstructures,
has been proved to cause the increases of shear viscosity~\cite{seto2013discontinuous}. 
This also explains the observation that 
the curves of the viscosity [\figrefp{fig:rheology}{a}] and particle contact number [\figrefp{fig:rotational_rearrangement}{b}]
have the similar shapes for the negative relative torques. 
However, when the relative torque is negative, 
it is seen that the shear viscosity decreases with the increases of $\phi$. 
Such an abnormal phenomenon, 
not seen for the positive torques,
indicates dependence of the viscous response on the torque (or shear) direction.


\Figrefp{fig:rheology}{c} shows the dependence
of the first normal stress coefficient $\Psi_1$ on $\tilde{T}_{\mathrm{A}}$,
which follows an imperfect odd function.
A close look at the curves around the origin can be found in \figref{fig_A10} of \Appendref{app:figures}.
The result reveals that the suspensions of particles self-rotating in the same direction,
i.e., chiral suspensions, 
are not the same as conventional complex fluids~\cite{mewis2012colloidal,dbouk2013normal,seto2018normal, larson1999structure}
or chiral viscous fluids~\cite{hargus2020time,han2021fluctuating},
but have the mixed feature of the two. 
From a physical scope, we explain that such an imperfect oddness has two origins,
i.e., odd responses of the self-rotating particles that break the parity and time-reversal symmetry of the suspensions,
and shear-induced microstructures.
Although mixed, 
the former origin is dominant over the latter one. 


On the other hand, 
from a mathematical scope,
we decompose the first normal stress coefficient into the odd (corresponding to the odd viscosity $\lambda_{\mathrm{odd}}$) 
and even ($\lambda_{\mathrm{even}}$) parts,
and plot them in terms of dimensionless shear rate 
$\tilde{\Gamma} \equiv \dot{\gamma}/(|\Vec{T}_{\mathrm{A}} \cdot \Vec{e}_{z}| / 6\pi \eta_0 a^3)$
for $\Vec{T}_{\mathrm{A}} \cdot \Vec{e}_{z} > 0$
[see Figs.\,\ref{fig:rheology}~(d) and (e)].
In the figures, we observe nonvanishing values of $\lambda_{\mathrm{odd}}$ and $\lambda_{\mathrm{even}}$.
In details, 
$|\lambda_{\mathrm{odd}}|$ decreases with $\tilde{\Gamma}$
but increases with $\phi$,
indicating the positive dependence of the odd viscosity on torque density.
For $\lambda_{\mathrm{even}}$, 
one can observe that its value transits from negative to positive as the shear rate increases.
Additionally, as the same feature of passive suspensions~\cite{seto2018normal}, 
increasing the areal fraction leads the value of $\lambda_{\mathrm{even}}$ to decrease first and then increase.


\section{Conclusions}
This paper reports by computational simulations that 
particles counterrotating with respect to the vorticity of imposed shear flows 
can self-organize into stripes. 
The mechanisms of such a phenomenon have been explained by giving insight into three collective particle behaviors:
shear-induced diffusion, rotation-induced rearrangement, and edge flows. 
The shear-induced diffusion leads to $y$-directional effective repulsion between the particles, 
whereas the rotation-induced rearrangement results in $y$-directional effective attraction for particles 
driven by the positive torques.
The stripelike aggregates only appear when the effective attraction is dominant. 
On the other hand, zero and reversed particle edge flows within the stripes guarantee the stability of the stripes. 
For the areal fraction $ 0.3 \le \phi \le 0.5$, 
the simulation result suggests that the average stripe width is independent of the areal fraction 
but linearly dependent on the relative torque intensity. 
By conducting a theoretical study, 
we also verify such a result from the scope of particle dynamics. 
Moreover, the rheological result suggests that 
the rotation of particles and formation of particle stripes 
lead to the decrease of shear and rotational viscosities.
The first normal stress coefficient exhibits asymmetric responses for opposite shears, 
suggesting the presence of odd viscosity in the suspension of self-rotating particles.
However, unlike chiral viscous fluids, 
the chiral suspension also exhibits some ``even viscosity'' responses.

The current paper does not take into account the thermal agitations or interparticle repulsion, 
which are sometimes essential for the collective behaviors of rotating active matters. 
To realize the above simulation results through experiments, 
we suggest producing the stable Taylor-Couette flows of dense suspension of self-rotating particles.
The particles should have a rough surface and negligible electrostatic interactions and deposit during the operation process.
To undergo the self-rotating, 
the particles could be functionalized and driven by rotating magnetic fields, light fields and other external resources. 
For the expectation, concentric particle rings will be formed in the middle height of the container.


\begin{acknowledgments}
We acknowledge Y. Hosaka and R. Podgornik for valuable discussion. Z.Z. thanks Z. Hou and the Postdoctor Association of WIUCAS for helpful discussions. The work was supported by the startup fund of Wenzhou Institute, University of Chinese Academy of Sciences (No.WIUCASQD2020002) and National Natural Science Foundation of China (No.12174390 and No.12150610463).
\end{acknowledgments}


\appendix

\section{Stress tensor for active suspensions}
\label{app:stress_tensor}
For the stress tensor, its original expression is given by 
\begin{align}
    \Tens{\sigma} &= 
    - p_0 \Tens{I}
    + 2 \eta_0 \Tens{E}^{\infty}
    - \frac{1}{V} \sum_{i > j}
    \Vec{r}_{ij} \Vec{F}_{\mathrm{Int},ij} \notag \\
&= 
    - p_0 \Tens{I}
    + 2 \eta_0 \Tens{E}^{\infty} \notag  \\
 &\quad   - \frac{1}{V} 
  \biggl(
    \mathrm{sym}
    \sum_{i > j}
    \Vec{r}_{ij} \Vec{F}_{\mathrm{Int},ij} 
    +  \mathrm{asym} 
    \sum_{i > j} 
    \Vec{r}_{ij} \Vec{F}_{\mathrm{Int},ij}
    \biggr),
\end{align}
where $\Tens{I}$ is an identity matrix,
$p_0$ represents the hydrostatic pressure (in our paper $p_0 = 0$),
$\Vec{F}_{\mathrm{Int},ij}$ is the interaction force vector acting on $i$,
and the symbols of $\mathrm{sym}$ and $\mathrm{asym}$ denote the symmetric and asymmetric parts of the tensor, respectively. 
Since 
\begin{align}
    \mathrm{asym}
    \sum_{i > j} 
    \Vec{r}_{ij} \Vec{F}_{\mathrm{Int},ij} 
    & = 
    \sum_{i > j} \Tens{\epsilon} \cdot 
    \Vec{r}_{ij}\times 
    \Vec{F}_{\mathrm{Int},ij} \\
    & =
    \sum_{i > j} \Tens{\epsilon} \cdot \Vec{T}_{\mathrm{Int},ij},
\end{align}
and within our consideration 
\begin{equation}
    \Vec{F}_{\mathrm{Int},ij} 
    = 
    \Vec{F}_{\mathrm{H},ij} 
    + 
    \Vec{F}_{\mathrm{C},ij}
\end{equation}
it is obtained 
\begin{equation}
\begin{split}
    \Tens{\sigma} &= 
    2 \eta_0 \Tens{E}^{\infty}
    - 
    \frac{1}{V} 
    \mathrm{sym}
    \sum_{i > j}
    \Vec{r}_{ij} 
    \left( \Vec{F}_{\mathrm{H},ij} 
    +
    \Vec{F}_{\mathrm{C},ij} \right) \\
&\quad  - \frac{1}{V} 
    \sum_{i > j}  \Tens{\epsilon} \cdot 
    \left( 
    \Vec{T}_{\mathrm{H},ij} 
    + 
    \Vec{T}_{\mathrm{C},ij} \right),
    \end{split}
\end{equation}
where $\Vec{T}_{\mathrm{Int},ij}$ represents the interaction torque vector from particles $j$ to $i$,
and the subscripts $\mathrm{H}$ 
and $\mathrm{C}$ indicate the forces/torques are due to the hydrodynamic lubrication and frictional contact, respectively. 
In the overdamped dynamics,
the total torques acting on respective particles satisfies
\begin{equation}
  \Vec{T}_{\mathrm{S},i}
  +
  \sum_{j \neq i} 
  \left( 
  \Vec{T}_{\mathrm{H},ij}
  +
  \Vec{T}_{\mathrm{C},ij}
  \right)
  +
  \Vec{T}_{\mathrm{A},i}
  =
  \Vec{0}.
\end{equation}
The stress tensor can be then rewritten as
\begin{equation}
\begin{split}
    \Tens{\sigma} &= 
    2 \eta_0 \Tens{E}^{\infty}
 - 
    \frac{1}{V}
    \mathrm{sym}
    \sum_{i > j}
    \Vec{r}_{ij} \left(
        \Vec{F}_{\mathrm{H},ij}
        +
        \Vec{F}_{\mathrm{C},ij}
        \right) \\
 &\quad  + \frac{1}{V} \sum_{i} \Tens{\epsilon} \cdot \left( \Vec{T}_{\mathrm{A},i} + \Vec{T}_{\mathrm{S},i} \right).
     \end{split}
\end{equation}
By substituting that $\Vec{T}_{\mathrm{A},i} = \Vec{T}_{\mathrm{A}}$ for the uniform active torque
and $\sum_{i} \Vec{T}_{\mathrm{S},i} = - 8 \pi \eta_0 a^3 N ( \langle \Vec{\Omega} \rangle - \Vec{\Omega}^{\infty} )$,
we finally obtain the expression shown in Eq.~\eqref{eq:stress_tensor}.


\section{Rotational viscosity}
\label{app:rotational_vis} 
Rotational viscosity is a rheological property that 
describes the damped transfer of angular momentum from the local element (internal) to the surroundings (external).
Such an element could be suspension, such as in chiral fluids, 
or a particle, such as in liquid crystals. 
According to the definition, the rotational viscosity, or vortex viscosity, 
can be obtained from the asymmetric component of stress tensor
\begin{equation}
  \mathrm{asym}(\Tens{\sigma}) = \eta_{\mathrm{rot}} \Tens{\epsilon} \cdot 
  \left(
  2 \Vec{\Omega}_{\mathrm{e}} - \Vec{\nabla} \times \Vec{U}^{\infty}
  \right),
  \label{eq:rotationalviscositydef}
\end{equation}
where $\Vec{\Omega}_{\mathrm{e}}$ represents the angular velocity of the local element 
(in our paper equal to $\langle \Vec{\Omega} \rangle$),
and the external vorticity $\Vec{\nabla} \times \Vec{U}^{\infty} = 2 \Vec{\Omega}^{\infty}$.
Applying Eq.~\eqref{eq:rotationalviscositydef} to our work, the rotational viscosity is in the form of
\begin{equation}
    \eta_{\mathrm{rot}} 
    =
    \frac{N}{2V} 
    \left(
    \frac{\Vec{T}_{\mathrm{A}} \cdot \Vec{e}_{z}} 
    {(\langle \Vec{\Omega} \rangle 
    - \Vec{\Omega}^{\infty}) \cdot \Vec{e}_{z}}
    -
    8 \pi \eta_0 a^3
    \right),
    \label{eq:rotationalviscosity1}
\end{equation}
where the total volume of the suspension $V = 2a A$ for a monolayer of particles, 
and $A$ is the area of the monolayer. 
Since the particle areal fraction $\phi = \pi a^2 N / A$, 
Eq.~\eqref{eq:rotationalviscosity1} can be rewritten by 
\begin{equation}
    \eta_{\mathrm{rot}} 
    = 
    2 \phi \eta_0
    \left(
    \frac{\Vec{T}_{\mathrm{A}} \cdot \Vec{e}_{z}} 
    { 8\pi \eta_0 a^3 (\langle \Vec{\Omega} \rangle 
    - \Vec{\Omega}^{\infty}) \cdot \Vec{e}_{z}} - 1
    \right).
    \label{eq:rotationalviscosity2}
\end{equation}
It should be noted that for the system consisting of discrete passive particles and solvent, 
the rotational friction coefficient, $8 \pi \eta_0 a^3$, 
plays a role to balance the angular momentum between the particles and the solvent. 
However, the rotational friction coefficient is an intrinsic property typical of particle-solvent systems. 
It cannot conserve the angular momentum when the discrete particles are actively rotating.
As a result, the rotational viscosity is necessary to be additionally considered. 
Equations~\eqref{eq:rotationalviscosity1} and~\eqref{eq:rotationalviscosity2} show the rotational viscosity that describes the nonhydrodynamically damped transfer of angular momentum from the particle to the surrounding fluid.
By introducing the nondimensionalized variables, 
we can finally obtain the expression shown in Eq.~\eqref{eq:rotationalviscosity}.


\section{Theoretical study of phase boundary%
\label{app:phase_boundary}
}
\begin{figure}[htb]
  \includegraphics[width=0.3\textwidth]{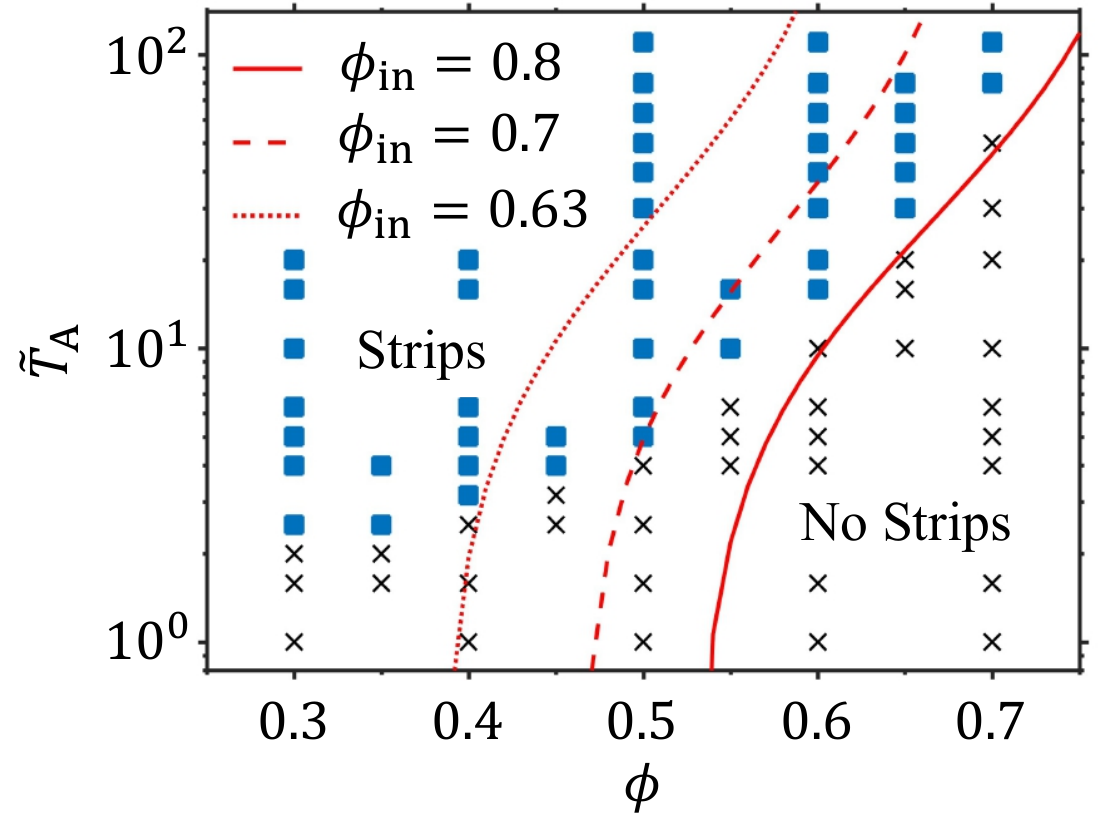}
    \caption{\label{fig_A1}
    Nonequilibrium phase diagram with various phase boundaries 
    that are predicted using various packing fractions of in-stripe particles.}
\end{figure}

Here we show the method of theoretical prediction of the phase boundary 
show in \figrefp{fig:characterization}{b}. 
Based on the conservation of total particle volume before and after applying active torques, one can write
\begin{equation}
  \phi \tilde{L}^2_{\mathrm{b}}
  =
  \phi_{\mathrm{in}} N_{\mathrm{s}} \tilde{L}_{\mathrm{s}} \tilde{L}_{\mathrm{b}},
  \label{eq:particleconservation}
\end{equation}
where $\tilde{L}_{\mathrm{b}}$ represents the radius-scaled side length of simulation box,
$\phi_{\mathrm{in}}$ the areal fraction of the particles in the stripes (or the packing fraction in the main text), 
and $N_{\mathrm{s}}$ the number of the particle stripes. 
For the particle configurations consisting of fully developed stripes, we also have 
\begin{equation}
  N_{\mathrm{s}} 
  \left( 
  \tilde{L}_{\mathrm{s}} + \tilde{L}_{\mathrm{g}} 
  \right) 
  =
  \tilde{L}_{\mathrm{b}}, 
  \label{eq:stripesum}
\end{equation}
where $\tilde{L}_{\mathrm{g}}$ is the average width of interstripe gaps. 
Combing Eqs.~\eqref{eq:particleconservation} and~\eqref{eq:stripesum} results in
\begin{equation}
  \tilde{L}_{\mathrm{s}} 
  = 
  \frac{\phi \tilde{L}_{\mathrm{g}}}{\phi_{\mathrm{in}} - \phi}. 
  \label{eq:phaseboundary}
\end{equation}
By substituting the linear equation of $\tilde{L}_{\mathrm{s}} \approx 0.22 \tilde{T}_{\mathrm{A}} + 3.92 $,
Eq.~\eqref{eq:phaseboundary} becomes
\begin{equation}
  \tilde{T}_{\mathrm{A}} 
  = 
  \frac{\phi \tilde{L}_{\mathrm{g}}}{0.22 
  \left( 
  \phi_{\mathrm{in}} - \phi
  \right) } - 17.82. 
  \label{eq:phaseboundaryfinal}
\end{equation}
To describe the phase boundary, 
we set $\tilde{L}_{\mathrm{g}}$ to the minimum critical value, i.e., $\tilde{L}_{\mathrm{g}} = 2$.
The phase boundary (solid red line) shown in Fig.\,1\,(b) of the main text is plotted by considering $\phi_{\mathrm{in}} = 0.8$. 
In \figref{fig_A1}, 
we also show the phase boundaries for various values of $\phi_{\mathrm{in}}$. 
When $\phi = 0.5$, 
we obtain from the simulation data that 
$\phi_{\mathrm{in}} \approx 0.7$ and $\tilde{L}_{\mathrm{g}} \approx 2$.
By substituting these values into Eq.~\eqref{eq:phaseboundaryfinal} and plotting it in the phase diagram, 
we find it successfully describes the phase boundary for $\phi = 0.5$. 
Similarly, when we substitute the simulation result of 
$\phi_{\mathrm{in}} \approx 0.63$ and $\tilde{L}_{\mathrm{g}} \approx 2.5$ for $\phi = 0.4$,
a good agreement is also achieved.

\section{\label{app:figures}Supplemental figures}

\begin{figure}[htb]
  \includegraphics[width=0.4\textwidth]{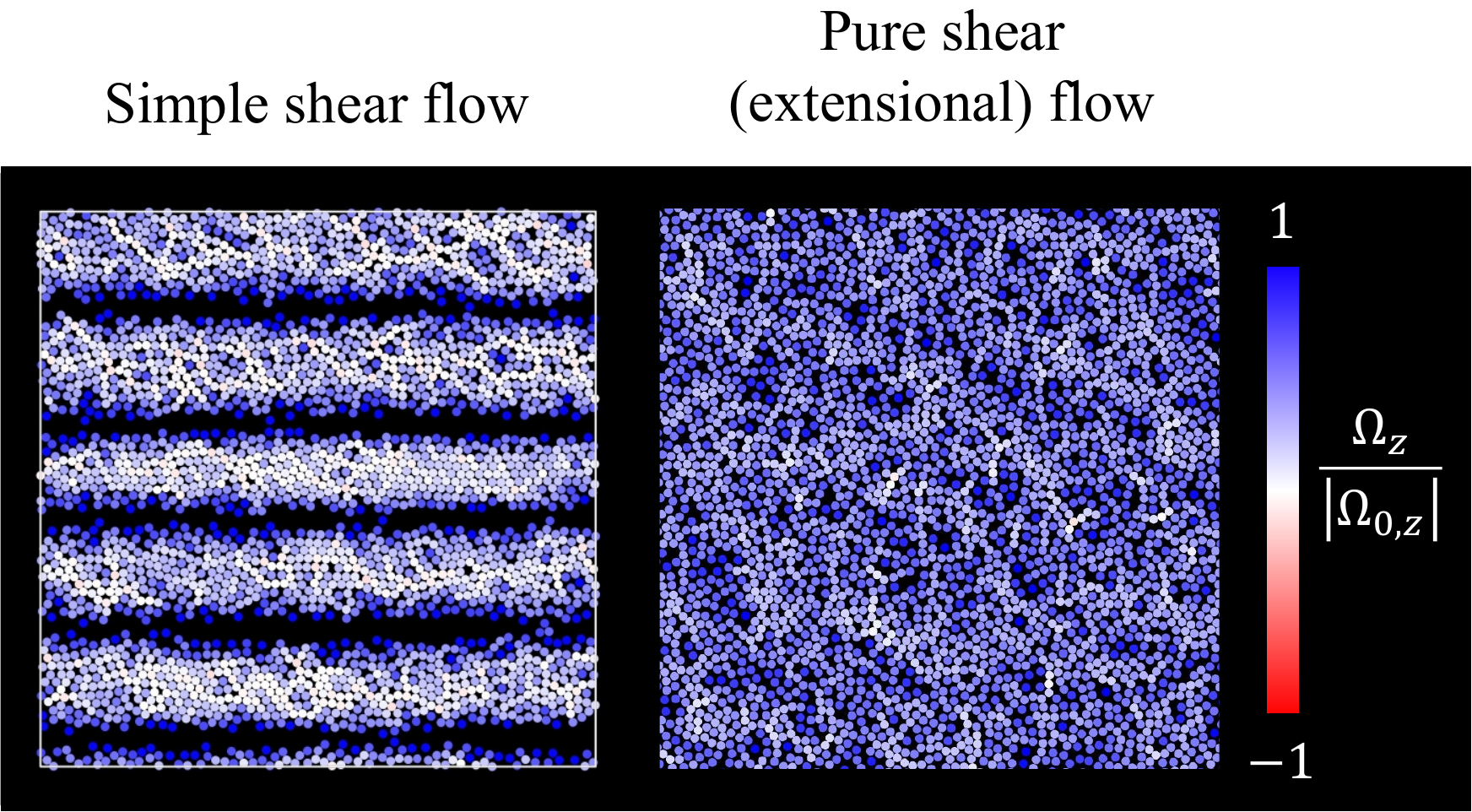}
  \caption{
  \label{fig_A2}
  Representative snapshots of fully developed particle configuration
for particle areal fraction $\phi = 0.6$, 
relative torque $\tilde{T}_{\mathrm{A}} = 60$, 
and in simple shear and pure shear flows. 
The color of the particle denotes the relative angular velocity of the particle 
to the angular velocity of a single particle under $\tilde{T}_{\mathrm{A}}$.}
\end{figure}

\Figref{fig_A2} shows 
the representative snapshots of fully developed particle configuration
for particle areal fraction $\phi = 0.6$, 
relative torque $\tilde{T}_{\mathrm{A}} = 60$, 
and in simple shear and pure shear flows. 
The color of the particle denotes the relative angular velocity of the particle 
to the angular velocity of a single particle under $\tilde{T}_{\mathrm{A}}$.
The result suggests that 
the self-rotating particles can form stripelike aggregates in the simple shear flow
but cannot in the pure shear (or extensional) flow. 

\begin{figure}[htbp]
  \includegraphics[width=0.47\textwidth]{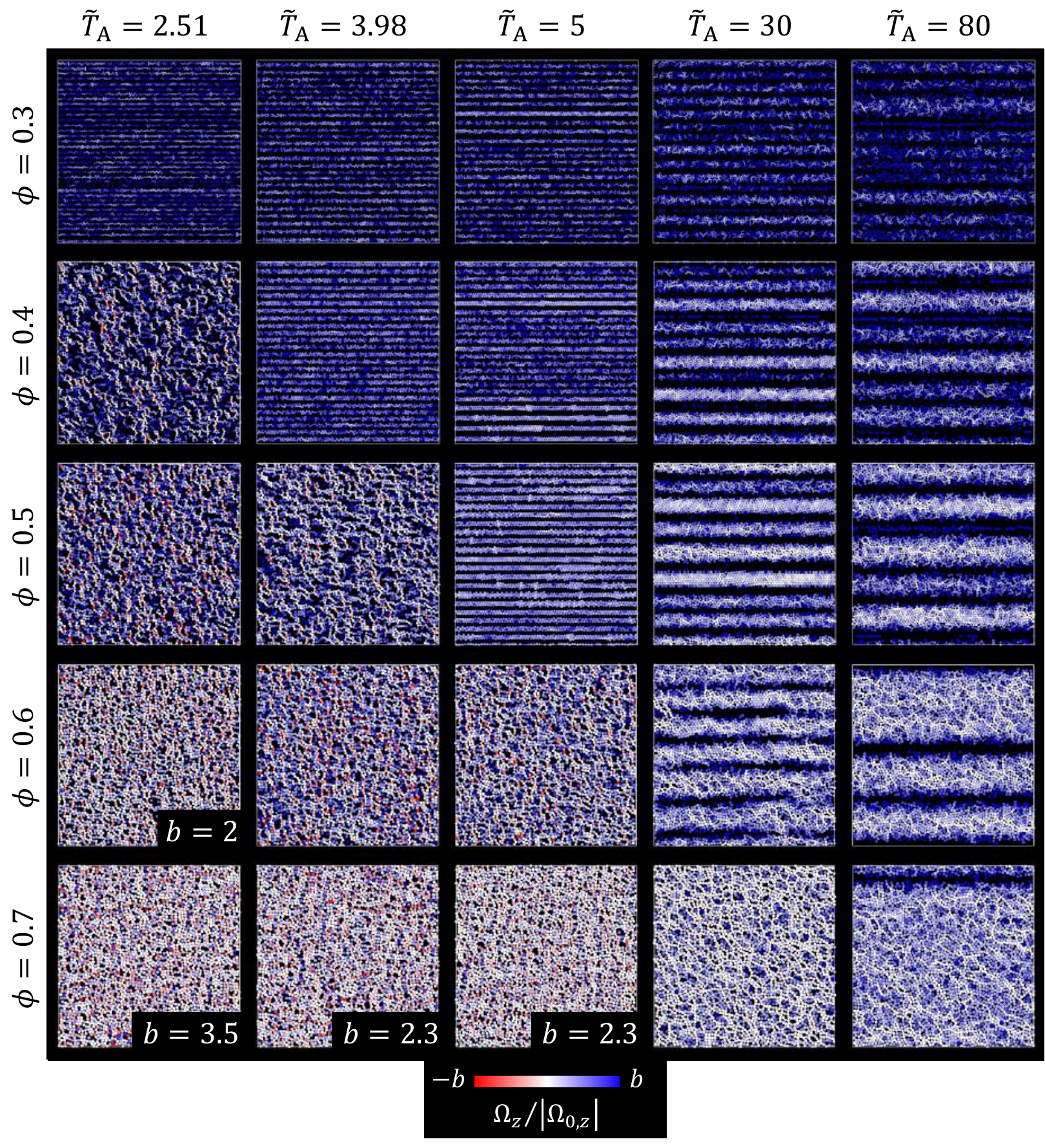}
  \caption{
    \label{fig_A3}
    Representative snapshots of particle configuration 
    for various positive relative torques and particle areal fractions. 
    The color of the particle denotes the relative angular velocity of the particle 
    to the angular velocity of a single particle under $\tilde{T}_{\mathrm{A}}$.
    In the color bar, $b = 1$ unless specified in the configurations.}
\end{figure}

\Figrefs{fig_A3}{fig_A4} show
the representative snapshots of particle configuration 
for various positive relative torques and particle areal fractions. 
The color of the particle denotes the relative angular velocity of the particle 
to the angular velocity of a single particle under $\tilde{T}_{\mathrm{A}}$.
In the color bar, $b = 1$ unless specified in the configurations.
In \figref{fig_A3}, it is observed that 
the particles self-organize into stripes under selected simulation conditions, 
and force chains may exist within the stripes. 
The threshold torque intensity to obtain the particle stripes increases with the particle areal fraction. 

\begin{figure}[htbp]
  \includegraphics[width=0.47\textwidth]{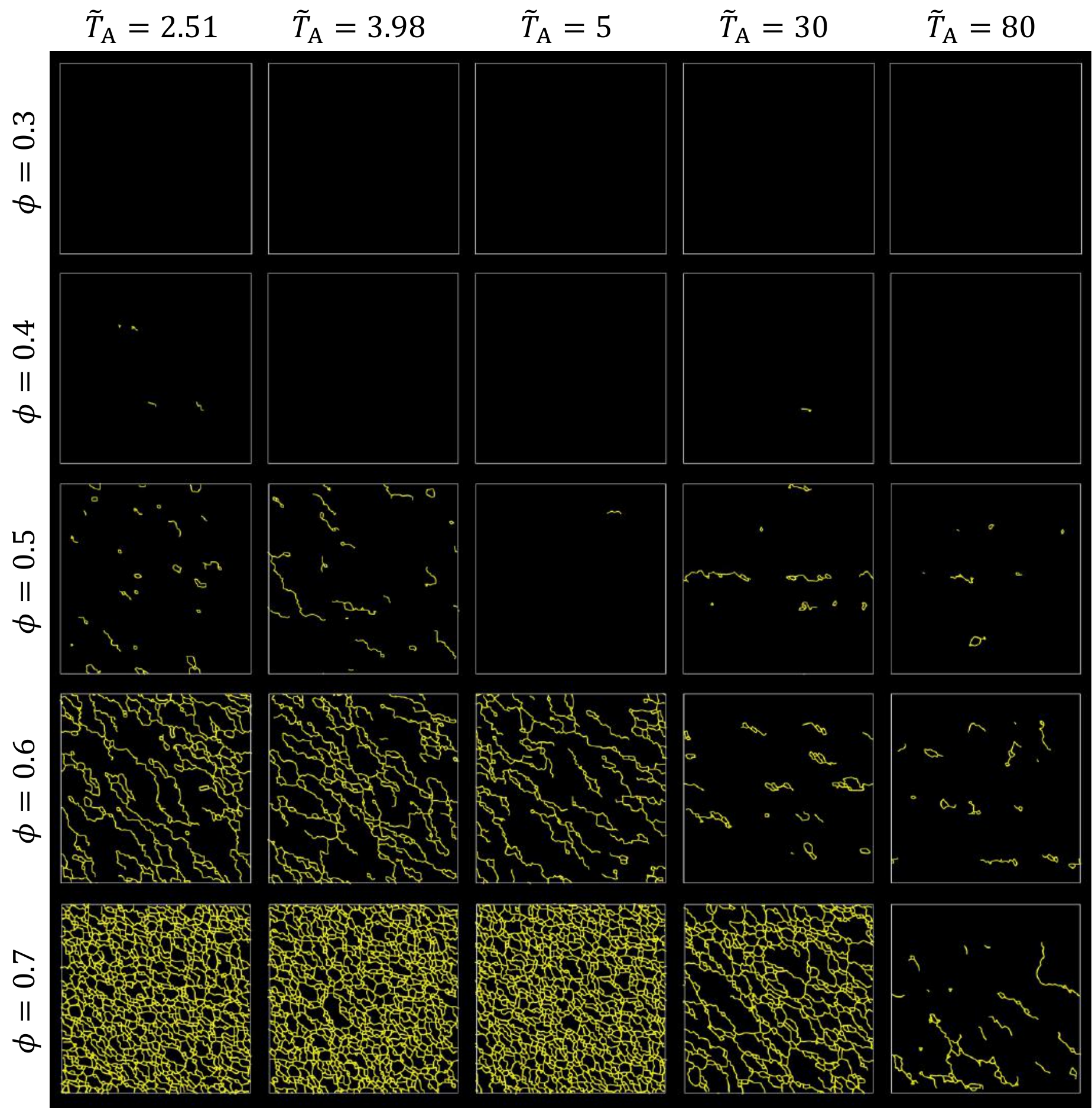}
  \caption{
    \label{fig_A4}
    Representative snapshots of force chain distribution 
    for various positive relative torques and particle areal fractions.}
\end{figure}

\begin{figure}[htbp]
  \includegraphics[width=0.47\textwidth]{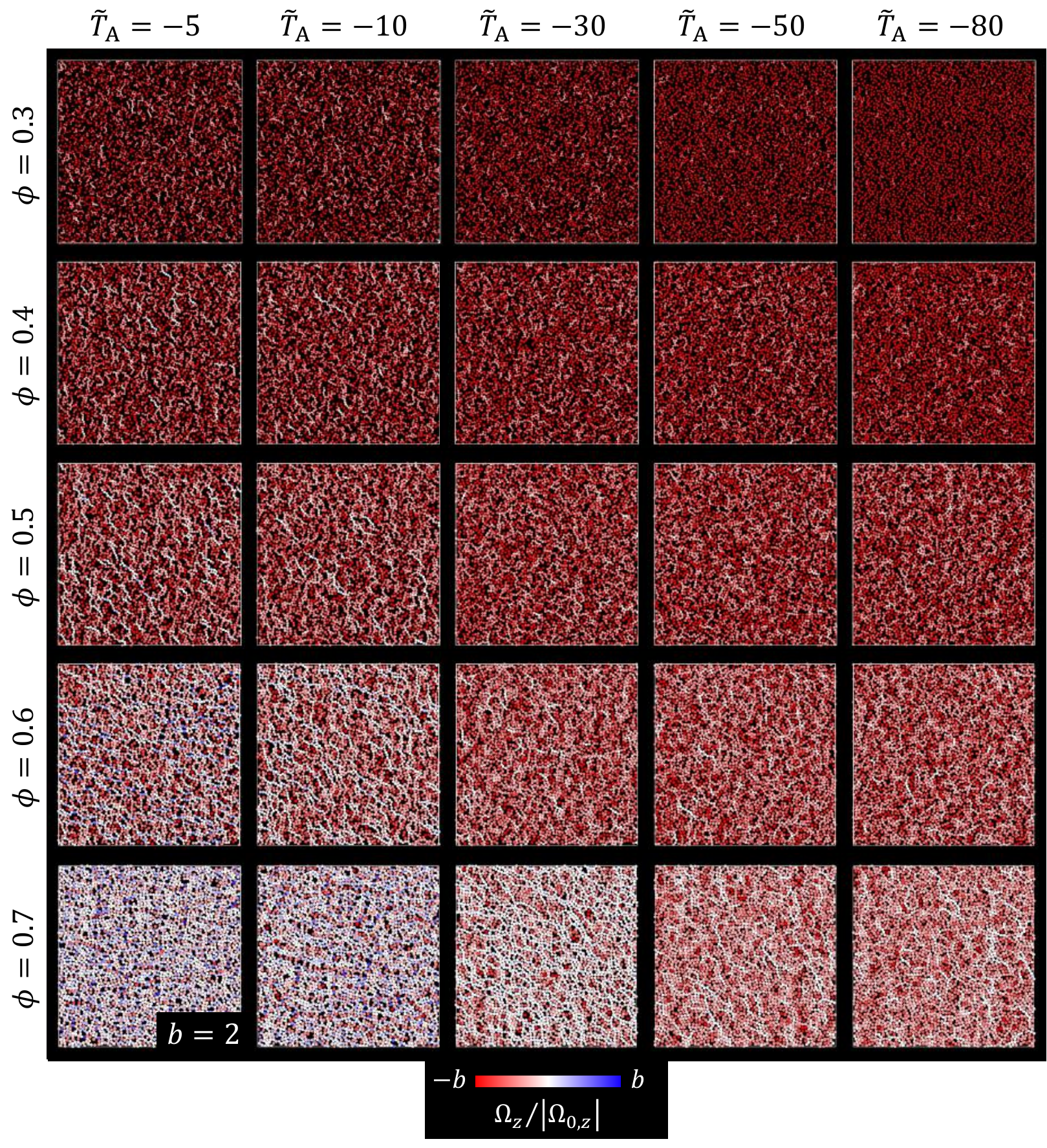}
  \caption{
    \label{fig_A5}
    Representative snapshots of particle configuration 
    for various negative relative torques and particle areal fractions. 
    The color of the particle denotes the relative angular velocity of the particle 
    to the angular velocity of a single particle under $\tilde{T}_{\mathrm{A}}$.
    In the color bar, $b = 1$ unless specified in the configurations.}
\end{figure}

\Figrefs{fig_A5}{fig_A6} show 
the representative snapshots of particle configuration 
for various negative relative torques and particle areal fractions. 
The color of the particle denotes the relative angular velocity of the particle 
to the angular velocity of a single particle under $\tilde{T}_{\mathrm{A}}$.
In the color bar, $b = 1$ unless specified in the configurations.
In \figref{fig_A5}, we observe that 
the particles do not formed stripes within a wide range of simulation parameters.
Besides, increasing the relative torque intensity leads the amount of force chains to decrease. 

\begin{figure}[htbp]
  \includegraphics[width=0.47\textwidth]{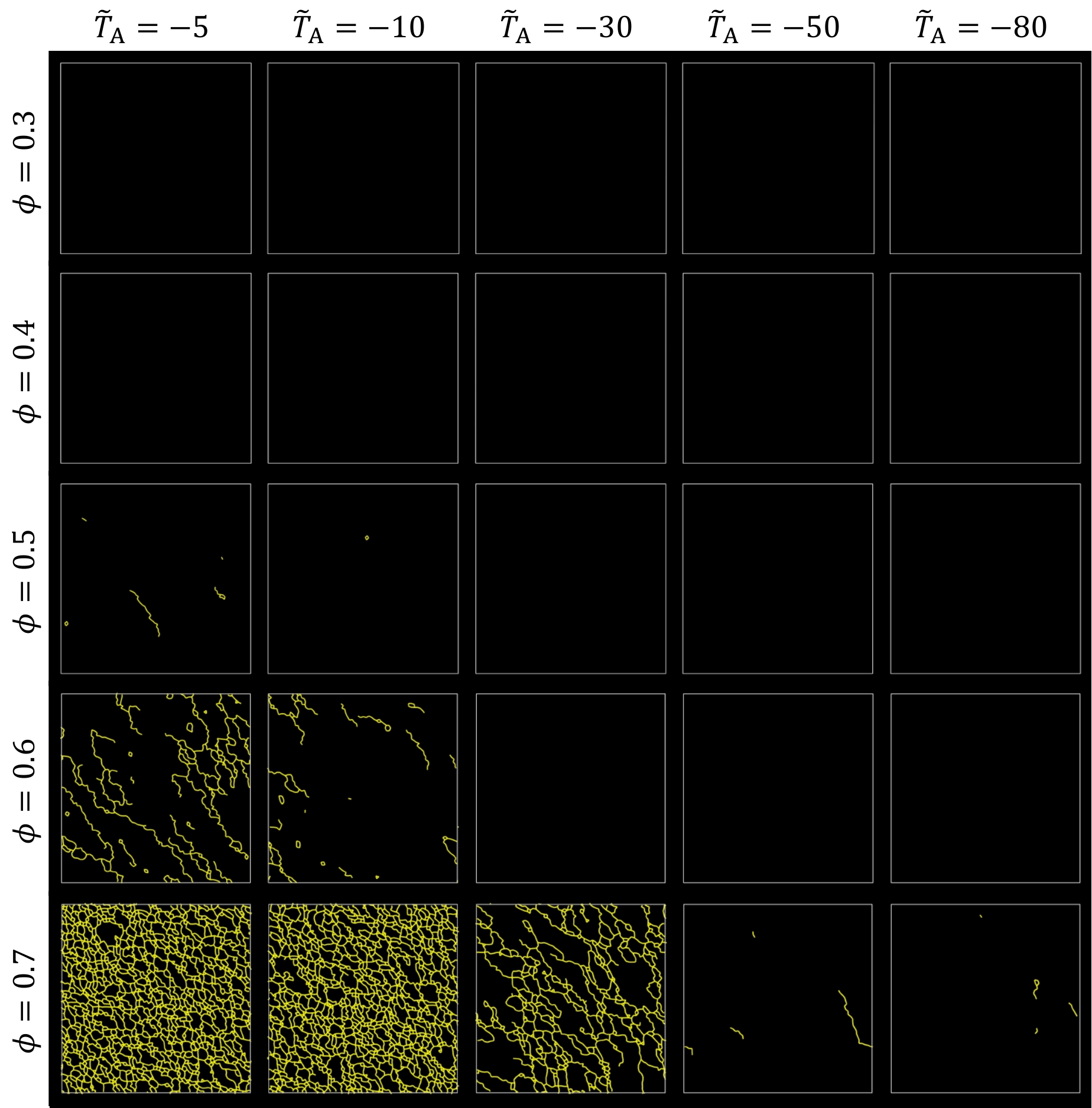}
  \caption{
    \label{fig_A6}
    Representative snapshots of force chain distribution 
    for various negative relative torques and particle areal fractions.}
\end{figure}

\begin{figure}[htbp]
  \includegraphics[width=0.47\textwidth]{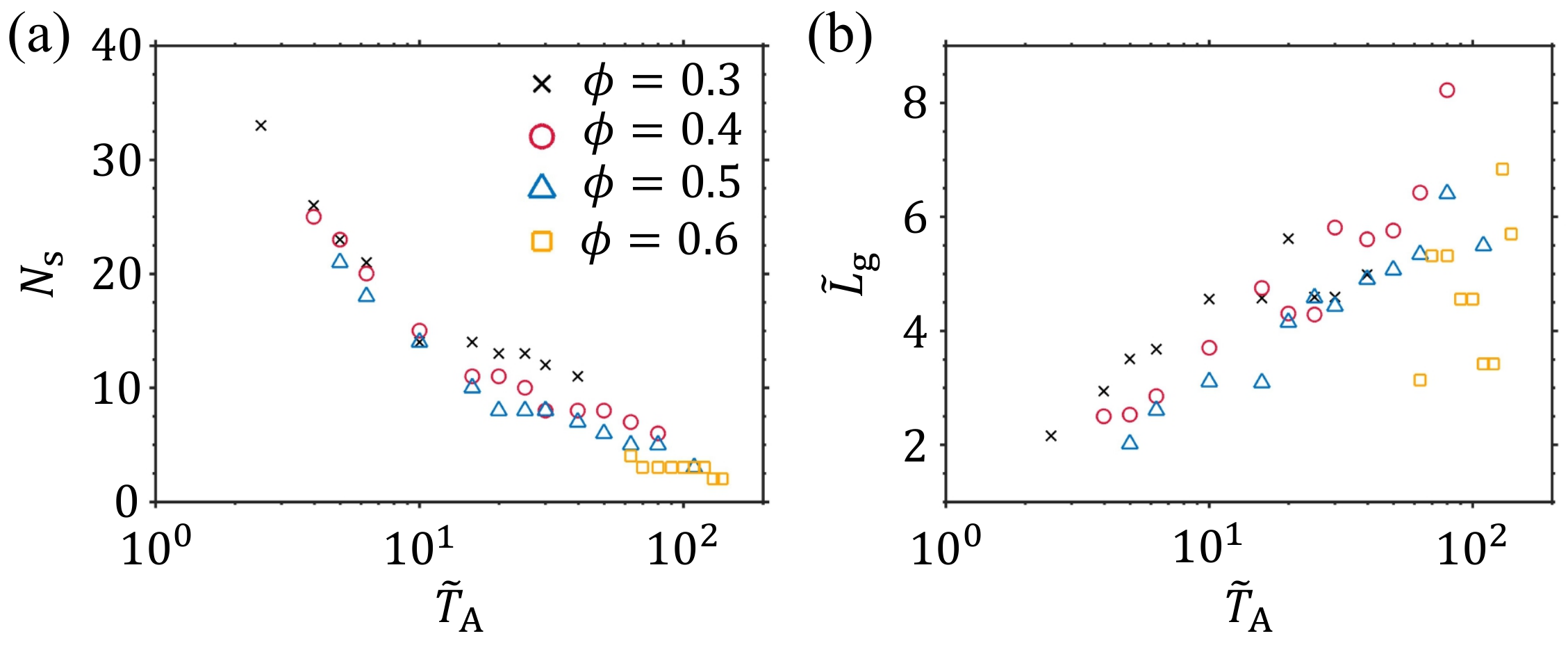}
  \caption{
    \label{fig_A7}
    (a) Number of particle stripes and (b) average width of interparticle gaps as a function of relative torque intensity, 
    for various particle areal fractions.}
\end{figure}

\Figref{fig_A7} shows 
(a) the number of particle stripes and (b) average width of interparticle gaps as a function of relative torque intensity, 
for various particle areal fractions.
In the figure, one can observe that 
increasing the value of $\tilde{T}_{\mathrm{A}}$ leads to the decrease of the stripe amount ($N_{\mathrm{S}}$)
and increase of the average gap width ($\tilde{L}_{\mathrm{g}}$).
However, large particle areal fractions give rise to the decrease of the stripe amount and average gap width. 

\begin{figure}[htbp]
  \includegraphics[width=0.47\textwidth]{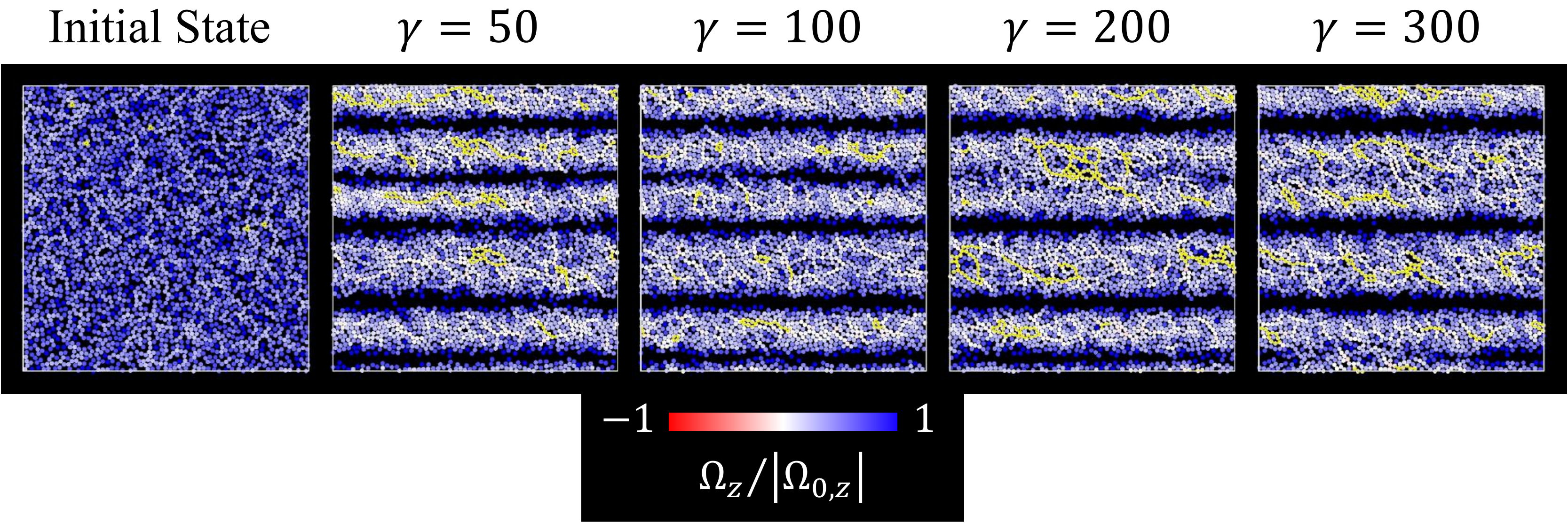}
  \caption{
    \label{fig_A8}
    Evolution of particle configuration 
    for the particle areal fraction $\phi = 0.6$ and relative torque intensity $\tilde{T}_{\mathrm{A}} = 50$. 
    The color of the particle denotes the relative angular velocity of the particle 
    to the angular velocity of a single particle under $\tilde{T}_{\mathrm{A}}$.
    The yellow bonds between the particles denote the interparticle contacts.}
\end{figure}

\Figref{fig_A8} shows
the evolution of particle configuration 
for the particle areal fraction $\phi = 0.6$ and relative torque intensity $\tilde{T}_{\mathrm{A}} = 50$. 
The color of the particle denotes the relative angular velocity of the particle 
to the angular velocity of a single particle under $\tilde{T}_{\mathrm{A}}$.
The yellow bonds between the particles denote the interparticle contacts.
Such a combination only happens when the interstripe gap is narrower than the particle diameter.

\begin{figure}[htbp]
  \includegraphics[width=0.47\textwidth]{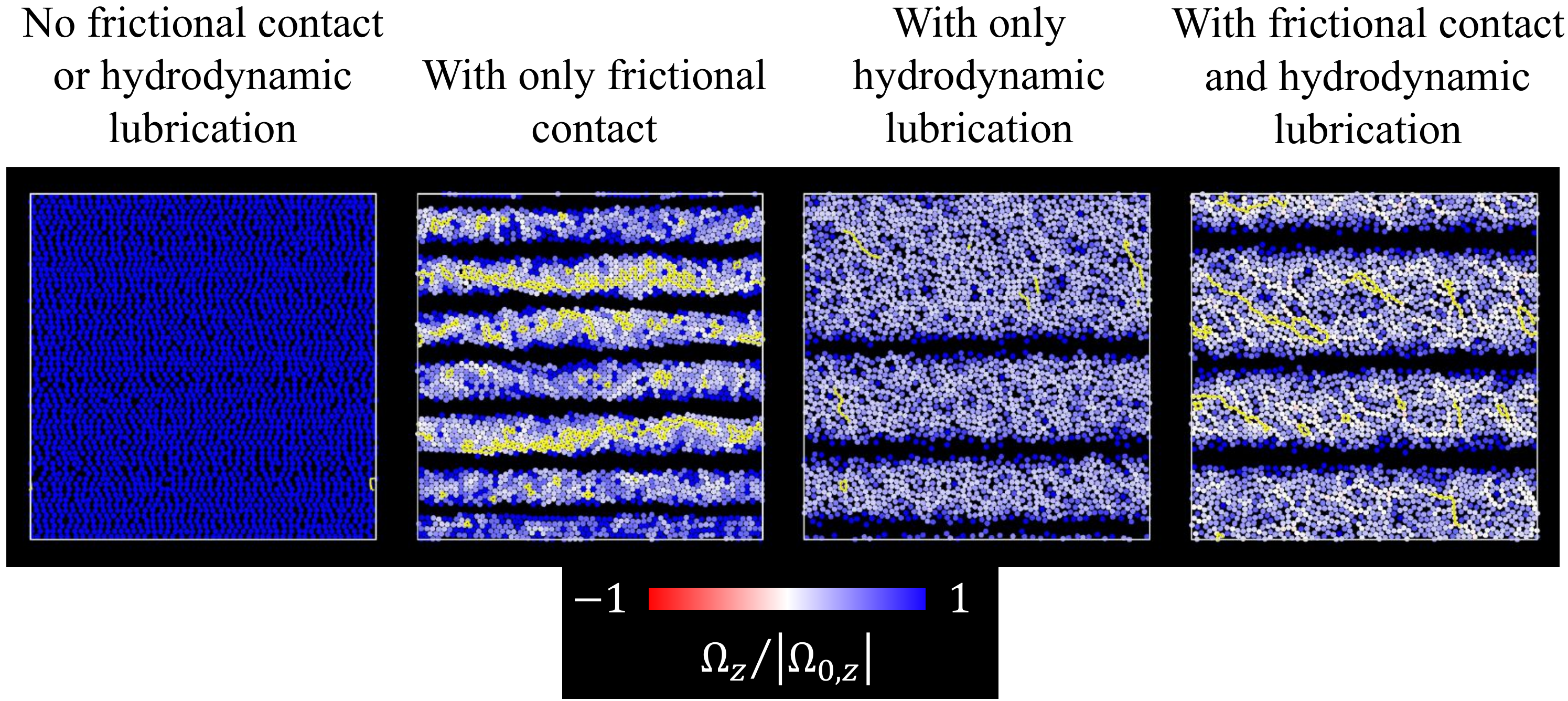}
  \caption{
    \label{fig_A9}
    Representative snapshots of particle configuration
    for the particle areal fraction $\phi = 0.6$,
    relative torque intensity $\tilde{T}_{\mathrm{A}} = 50$,
    and various interaction conditions.
    The color of the particle denotes the relative angular velocity of the particle 
    to the angular velocity of a single particle under $\tilde{T}_{\mathrm{A}}$.
    The yellow bonds between the particles denote the interparticle contacts.}
\end{figure}

\Figref{fig_A9} shows
the representative snapshots of particle configuration
for the particle areal fraction $\phi = 0.6$,
relative torque intensity $\tilde{T}_{\mathrm{A}} = 50$,
and various interaction conditions.
The color of the particle denotes the relative angular velocity of the particle 
to the angular velocity of a single particle under $\tilde{T}_{\mathrm{A}}$.
The yellow bonds between the particles denote the interparticle contacts.
We note that for the case without frictional contact or hydrodynamic lubrication, 
the only interparticle interaction is the excluded-volume interaction,
which in our paper is treated as the normal contact force.
For the case with only frictional contact,
we additionally take into account the tangential contact force between the particles. 
In the figure, it is observed that the particles can self-organize into stripelike aggregates, 
except for the case with only excluded-volume interaction. 
Additionally, as compared with the stripes formed by the particles with only hydrodynamic lubrication,
those with only frictional contact give rise to more uniform and hexagonal structures. 
When both frictional contact and hydrodynamic lubrication are taken into account,
the particle stripes show the mixture of features of the stripes with one of the two interactions.

\begin{figure}[htbp]
  \includegraphics[width=0.47\textwidth]{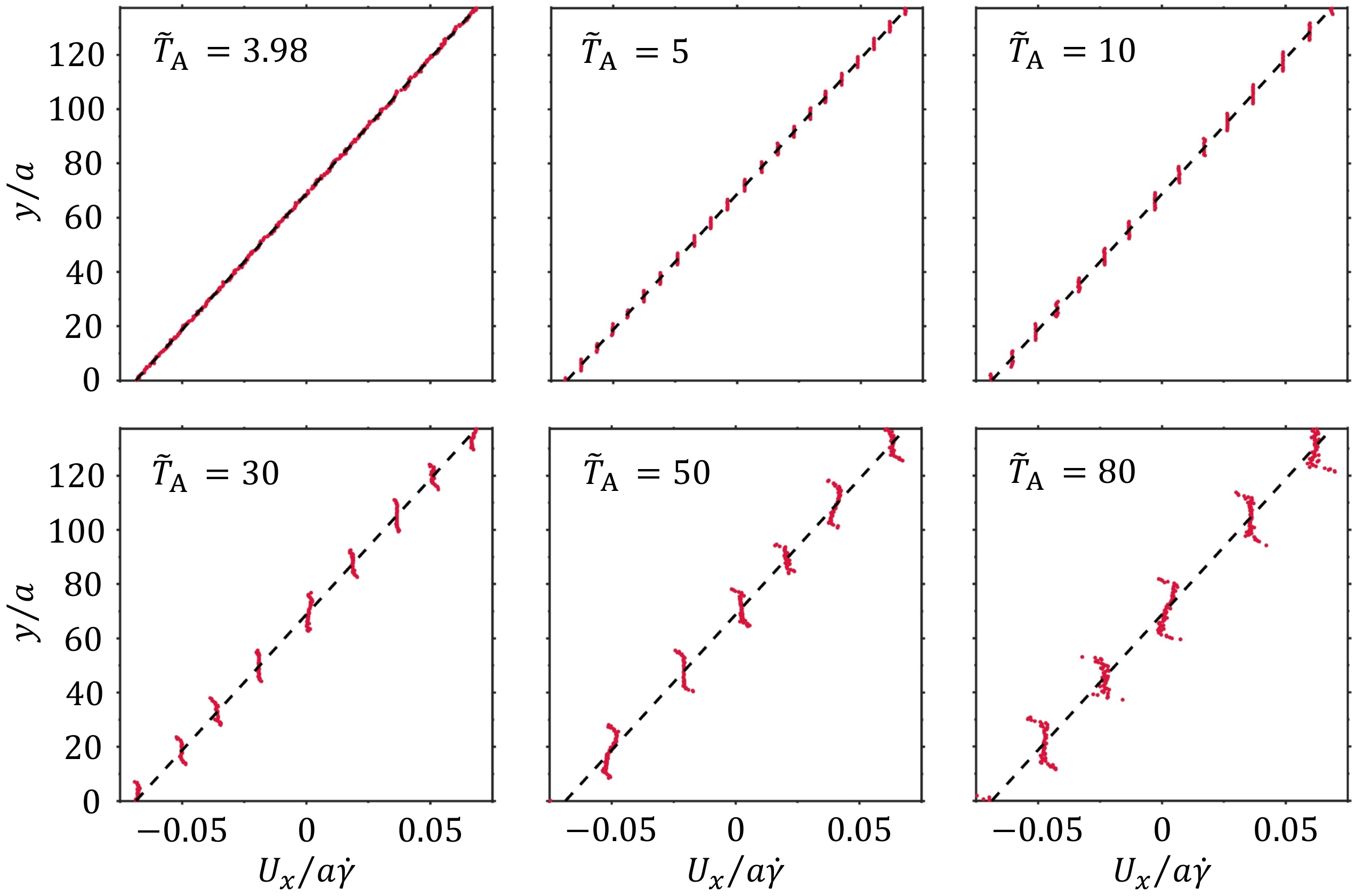}
  \caption{
    \label{fig_A10}
    Profiles of $x$ component velocity of the in-stripe particles, 
    for the particle areal fraction $\phi = 0.5$ and various relative torque intensities.}
\end{figure}

\Figref{fig_A10} shows 
the profiles of $x$ component velocity of the in-stripe particles, 
for the particle areal fraction $\phi = 0.5$ and various relative torque intensities.
In the figure, we observe that for $\tilde{T}_{\mathrm{A}} = 3.98$, 
no particle stripes are constructed 
and the particles translate in the same velocity with the imposed simple shear flow. 
However, for $\tilde{T}_{\mathrm{A}} = 5$ and $10$, 
narrow stripes emerge and translate as rigid bodies, 
with the velocity equal to the flow velocity at the half height of the stripe. 
A further increase of the relative torque intensity 
then results in the emergence of significant edge flows of the particles,
which are against the imposed flow. 

\begin{figure}[htbp]
  \includegraphics[width=0.3\textwidth]{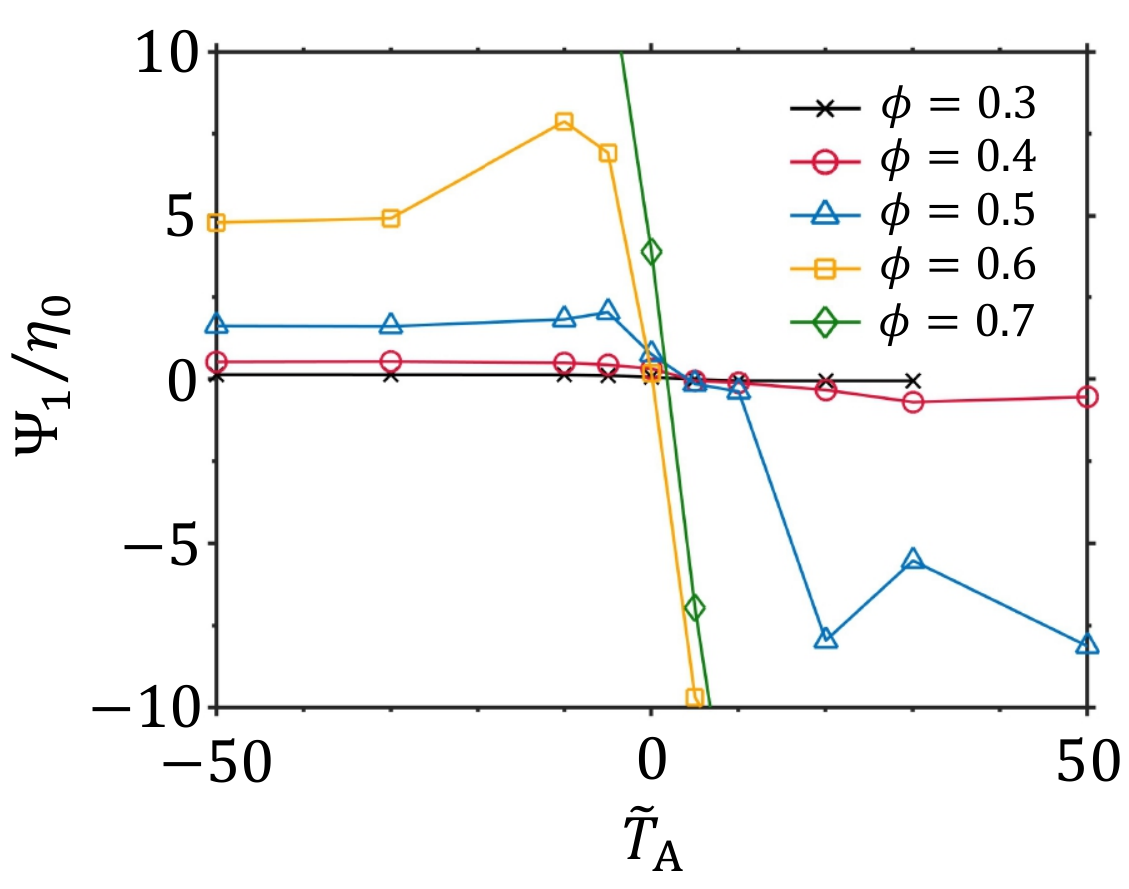}
  \caption{
    \label{fig_A11}
    Relative first normal stress coefficient as a function of relative torque intensity,
    for various particle areal fractions and in a zoomed-in window.}
\end{figure}

\Figref{fig_A11} shows 
the relative first normal stress coefficient as a function of relative torque intensity,
for various particle areal fractions and in a zoomed-in window.
In the figure, the curves show the slightly shifted odd shapes,
with the positive values at zero relative torque.

\nocite{*}

\providecommand{\noopsort}[1]{}\providecommand{\singleletter}[1]{#1}%
%


\end{document}